\documentclass{aa}

\usepackage{graphicx}

\usepackage[colorlinks]{hyperref}
\usepackage[all]{hypcap} 

\usepackage{listings}

\usepackage{epic,eepic}

\usepackage{color}
\usepackage{xspace}

\usepackage{cmbright}

\def\AAmarker#1{{#1}}

\hypersetup{
  linkcolor  = blue,
  citecolor  = blue,
  urlcolor   = blue,
  colorlinks = true,
}

\graphicspath{{./}}

\newcommand{\bea}{\begin{eqnarray}}
\newcommand{\eea}{\end{eqnarray}}

\def\b{\beta}

\def\K{{\rm K}\xspace}
\def\logg{{\log(g)}}

\def\rout{\ifmmode{r_{\rm out}}\else\hbox{$r_{\rm out}$}\fi}
\def\tmax{\ifmmode{\tau_{\rm max}}\else\hbox{$\tau_{\rm max}$}\fi}
\def\tstd{\ifmmode{\tau_{\rm std}}\else\hbox{$\tau_{\rm std}$}\fi}
\def\vmax{\ifmmode{v_{\rm max}}\else\hbox{$v_{\rm max}$}\fi}
\def\muE{\ifmmode{\mu_{\rm E}}\else\hbox{$\mu_{\rm E}$}\fi} 
\def\pE{\ifmmode{p_{\rm E}}\else\hbox{$p_{\rm E}$}\fi} 
\def\bmax{\ifmmode{\b_{\rm max}}\else\hbox{$\b_{\rm max}$}\fi}

\def\ang{\hbox{\AA}}
\def\Msun{\hbox{$\,$M$_\odot$} }

\def\Teff{\hbox{$\,T_{\rm eff}$} }
\def\alog#1{\times 10^{#1}}

\def\rout{\hbox{$r_{\rm out}$} }

\def\lstar{\ifmmode{\Lambda^*}\else\hbox{$\Lambda^*$}\fi} 
\def\Lstar{\ifmmode{\Lambda^*}\else\hbox{$\Lambda^*$}\fi} 
\def\Rop{\ifmmode{[R_{ij}]}\else\hbox{$[R_{ij}]$}\fi}

\def\Rji{\ifmmode{[R_{ji}]}\else\hbox{$[R_{ji}]$}\fi}
\def\Rstar{\ifmmode{[R_{ij}^*]}\else\hbox{$[R_{ij}^*]$}\fi}

\def\Rjistar{\ifmmode{[R_{ji}^*]}\else\hbox{$[R_{ji}^*]$}\fi}
\def\DRji{\ifmmode{[\Delta R_{ji}]}\else\hbox{$[\Delta R_{ji}]$}\fi}
\def\DRij{\ifmmode{[\Delta R_{ij}]}\else\hbox{$[\Delta R_{ij}]$}\fi}

\def\ns{\ifmmode{N_{\rm s}}          % Anzahl der tau-punkte
        \else\hbox{$N_{\rm s}$}\fi}
\def\mat#1{{\bf #1}}     % Macro fr Matrizen
\def\vek#1{{#1}}         % Macro fr Vektoren

\newcount\eqcount
\def
  \nummer{
    \global\advance\eqcount by 1
    (\the\eqcount)
  }

\def
  \numadv{
    \global\advance\eqcount by 1
  }

\def
   \numout#1{
     (\the\eqcount #1)
  }

\def\ivek#1#2{\ifmmode{\vek{I}^{#1}_{#2}}
        \else\hbox{$\vek{I}^{#1}_{#2}$}\fi}

\def\tmat#1#2{\ifmmode{\mat{t}^{#1}_{#2}}
        \else\hbox{$\mat{t}^{#1}_{#2}$}\fi}
\def\rmat#1#2{\ifmmode{\mat{r}^{#1}_{#2}}
        \else\hbox{$\mat{r}^{#1}_{#2}$}\fi}
\def\bvek#1#2{\ifmmode{\beta^{#1}_{#2}}
        \else\hbox{$\beta^{#1}_{#2}$}\fi}

\def\lp{\ifmmode{\lambda^+_\tau}           % lambda +
        \else\hbox{$\lambda^+_\tau$}\fi}
\def\lm{\ifmmode\lambda^-_\tau             % lambda -
        \else\hbox{$\lambda^-_\tau$}\fi}

\makeatletter

\newcommand\Autoref[1]{\@first@ref#1,@}
\def\@throw@dot#1.#2@{#1}% discard everything after the dot
\def\@set@refname#1{%    % set \@refname to autoefname+s using \getrefbykeydefault
    \edef\@tmp{\getrefbykeydefault{#1}{anchor}{}}%
    \def\@refname{\@nameuse{\expandafter\@throw@dot\@tmp.@autorefname}s}%
}
\def\@first@ref#1,#2{%
  \ifx#2@\autoref{#1}\let\@nextref\@gobble% only one ref, revert to normal \autoref
  \else%
    \@set@refname{#1}%  set \@refname to autoref name
    \@refname~\ref{#1}% add autoefname and first reference
    \let\@nextref\@next@ref% push processing to \@next@ref
  \fi%
  \@nextref#2%
}
\def\@next@ref#1,#2{%
   \ifx#2@ and~\ref{#1}\let\@nextref\@gobble% at end: print and+\ref and stop
   \else, \ref{#1}% print  ,+\ref and continue
   \fi%
   \@nextref#2%
}

\makeatother

\def\aasref@jnl#1{{\rm #1}}

\def\icarus{\aasref@jnl{ICARUS}}                   % Astronomical Journal
\def\aj{\aasref@jnl{AJ}}                   % Astronomical Journal
\def\araa{\aasref@jnl{ARA\&A}}             % Annual Review of Astron and Astrophys
\def\apj{\aasref@jnl{ApJ}}                 % Astrophysical Journal
\def\apjl{\aasref@jnl{ApJ}}                % Astrophysical Journal, Letters
\def\apjs{\aasref@jnl{ApJS}}               % Astrophysical Journal, Supplement
\def\ao{\aasref@jnl{Appl.~Opt.}}           % Applied Optics
\def\apss{\aasref@jnl{Ap\&SS}}             % Astrophysics and Space Science
\def\aap{\aasref@jnl{A\&A}}                % Astronomy and Astrophysics
\def\aapr{\aasref@jnl{A\&A~Rev.}}          % Astronomy and Astrophysics Reviews
\def\aaps{\aasref@jnl{A\&AS}}              % Astronomy and Astrophysics, Supplement
\def\azh{\aasref@jnl{AZh}}                 % Astronomicheskii Zhurnal
\def\baas{\aasref@jnl{BAAS}}               % Bulletin of the AAS
\def\jrasc{\aasref@jnl{JRASC}}             % Journal of the RAS of Canada
\def\memras{\aasref@jnl{MmRAS}}            % Memoirs of the RAS
\def\mnras{\aasref@jnl{MNRAS}}             % Monthly Notices of the RAS
\def\pra{\aasref@jnl{Phys.~Rev.~A}}        % Physical Review A: General Physics
\def\prb{\aasref@jnl{Phys.~Rev.~B}}        % Physical Review B: Solid State
\def\prc{\aasref@jnl{Phys.~Rev.~C}}        % Physical Review C
\def\prd{\aasref@jnl{Phys.~Rev.~D}}        % Physical Review D
\def\pre{\aasref@jnl{Phys.~Rev.~E}}        % Physical Review E
\def\prl{\aasref@jnl{Phys.~Rev.~Lett.}}    % Physical Review Letters
\def\pasp{\aasref@jnl{PASP}}               % Publications of the ASP
\def\pasj{\aasref@jnl{PASJ}}               % Publications of the ASJ
\def\qjras{\aasref@jnl{QJRAS}}             % Quarterly Journal of the RAS
\def\skytel{\aasref@jnl{S\&T}}             % Sky and Telescope
\def\solphys{\aasref@jnl{Sol.~Phys.}}      % Solar Physics
\def\sovast{\aasref@jnl{Soviet~Ast.}}      % Soviet Astronomy
\def\ssr{\aasref@jnl{Space~Sci.~Rev.}}     % Space Science Reviews
\def\zap{\aasref@jnl{ZAp}}                 % Zeitschrift fuer Astrophysik
\def\nat{\aasref@jnl{Nature}}              % Nature
\def\iaucirc{\aasref@jnl{IAU~Circ.}}       % IAU Cirulars
\def\aplett{\aasref@jnl{Astrophys.~Lett.}} % Astrophysics Letters
\def\apspr{\aasref@jnl{Astrophys.~Space~Phys.~Res.}}
\def\bain{\aasref@jnl{Bull.~Astron.~Inst.~Netherlands}} 
\def\fcp{\aasref@jnl{Fund.~Cosmic~Phys.}}  % Fundamental Cosmic Physics
\def\gca{\aasref@jnl{Geochim.~Cosmochim.~Acta}}   % Geochimica Cosmochimica Acta
\def\grl{\aasref@jnl{Geophys.~Res.~Lett.}} % Geophysics Research Letters
\def\jcp{\aasref@jnl{J.~Chem.~Phys.}}      % Journal of Chemical Physics
\def\jgr{\aasref@jnl{J.~Geophys.~Res.}}    % Journal of Geophysics Research
\def\jqsrt{\aasref@jnl{J.~Quant.~Spec.~Radiat.~Transf.}}
\def\memsai{\aasref@jnl{Mem.~Soc.~Astron.~Italiana}}
\def\nphysa{\aasref@jnl{Nucl.~Phys.~A}}   % Nuclear Physics A
\def\physrep{\aasref@jnl{Phys.~Rep.}}   % Physics Reports
\def\physscr{\aasref@jnl{Phys.~Scr}}   % Physica Scripta
\def\planss{\aasref@jnl{Planet.~Space~Sci.}}   % Planetary Space Science
\def\procspie{\aasref@jnl{Proc.~SPIE}}   % Proceedings of the SPIE

\def\phxO{{\tt PHOENIX/1D}\xspace}

\newcommand{\nmodel}{37438\xspace}

\addto\extrasenglish{ 	
  
   }  

\begin{document}

\title{\AAmarker{The} NewEra model grid }

\titlerunning{NewEra model grid}
\authorrunning{Hauschildt et al.}
\author{Peter H. Hauschildt\inst{1}, T.~Barman\inst{2}, E.~Baron\inst{3,1,4},  J.P.~Aufdenberg\inst{5}, and A. Schweitzer\inst{1}}
\institute{
Hamburger Sternwarte, Gojenbergsweg 112, 21029 Hamburg, Germany;
\email{yeti@hs.uni-hamburg.de}
\and
Lunar and Planetary Laboratory, University of Arizona, Tucson, AZ 85721 USA; 
\and
Planetary Science Institute, 1700 E. Fort Lowell Rd, Ste 106, Tucson, AZ 85719 USA; 
\and
Homer L.~Dodge Dept.~of Physics and Astronomy, University of
Oklahoma, 440 W.  Brooks, Rm 100, Norman, OK 73019 USA
\and
Department of Physical Sciences, College of Arts and Sciences, Embry-Riddle Aeronautical University,
1 Aerospace Boulevard, Daytona Beach, FL 32114 USA
}

\date{Received date \ Accepted date}

\abstract
{Analyses of stellar spectra, stellar populations, and transit light
curves rely on grids of synthetic spectra and center-to-limb
variations (limb darkening) from model stellar atmospheres.  Extensive
model grids from PHOENIX, a generalized non-LTE 1D and 3D stellar
atmosphere code, have found widespread use in the astronomical
community, however current PHOENIX/1D models have been substantially
improved over the last decade.}
{To make these improvements available to
the community, we have constructed the NewEra LTE model grid
consisting of \nmodel models with $2300\K \le \Teff \le 12000\K$,
$0.0\le \logg \le 6.0$ metallicities [M/H] from $-4.0$ to $+0.5$, and for
metallicities $-2.0 \le [{\rm M/H}] \le 0.0$ additional $\alpha$ element
variations from 
$-0.2 \le [\alpha/{\rm Fe}] \le +1.2$ are included.}
{The models use databases of 851 million atomic
lines and 834 billion molecular lines and employ the Astrophysical
Chemical Equilibrium Solver for the equation of state.\AAmarker{All models in the NewEra grid
have been calculated in spherical symmetry because center-to-limb
variation differences from plane-parallel models are quite large for
giants and not insignificant for dwarfs.}}
{ All model data are provided in the Hierarchical Data Format 5 (HDF5) format,
including low and high sampling rate spectra.  These files also include a
variety of details about the models, such as the exact abundances and isotopic
patterns used and results of the atomic and molecular line selection. }
{\AAmarker{Although the model structures have small differences with  the previous
grid generation, the spectra show significant differences, mostly due to the updates
of the molecular line lists.}}

\keywords{Stars: atmospheres}

\maketitle

\section{Introduction}

\texttt{PHOENIX} \citep{parapap,jcam}, currently at version 20,  is a generalized NLTE 1D and 3D stellar 
atmosphere code that has been used
for a variety of astrophysical sources including stars \citep{2005A&A...436..677F,2010A&A...511A..83F,2018A&A...615A...6P},
irradiated planets \citep{2002ApJ...569L..51B,2008PhST..130a4033H,2019ApJ...871..235P},  novae \citep{2001ApJ...547.1057S}
supernovae
\citep{2002ApJ...574..293M,2003ApJ...586.1199B,2017MNRAS.467.2392F,2020ApJ...901...86D},
and AGN \citep{Casebeer:2008}.
\phxO\ has been used to compute grids of stellar atmosphere models \citep{2000ApJ...537..946B,MDpap,ng-giants,ng-hot,Husser_etal_2013} 
for modeling and analysis of stellar spectra. These model grids have 
found widespread use in the astronomical community \citep[for example,][]{Smitha:2025,Madore:2025,Savel:2025,Barnes:2024,Passegger:2019,Delgado:2016,Shporer:2014}; however, \phxO\  models have
substantially improved over the last decade.

Therefore, we present here the NewEra LTE grid of \AAmarker{spherically symmetric} stellar models, computed with 
the latest version of \phxO\ and the corresponding set of input physical data.
An NLTE version of the grid is in preparation and will be
discussed in a subsequent paper. As for the previous model grids, the purpose of
this grid is to aid in the analyses of stellar spectra, modeling stellar populations, and applications 
such as transit light curve modeling. The results are available online \AAmarker{(see section \ref{sec:software})} in detailed 
and compact GAIA DR4 formats for different use cases. The NewEra grid covers a similar
parameter range as the ACES grid \citep{Husser_etal_2013} and follows the same mass--$\Teff$--$\logg$
relation.

\section{Equation of state: ACES}

The Astrophysical Chemical Equilibrium Solver \citep[ACES,][]{Barman_2011}
equation of state  has been used  in PHOENIX beginning with
version 16 \AAmarker{and the ACES grid \citep{Husser_etal_2013}}. 
ACES  is a state-of-the-art treatment of the chemical
equilibrium in a stellar atmosphere. The method is based upon that of 
\citet{Smith_Missen_1982} incorporating new experimental and theoretical
thermodynamic data \citep{Barman_2011} for 839 species (84 elements,
289 ions, 249 molecules, 217 condensates) \AAmarker{with all available data updates applied}.
Chemical equilibrium concentrations for all  atomic
and molecular species (the species are selected at run time) are
determined by the pressure, temperature, and density. This calculation 
is repeated at each layer in the atmosphere for each model iteration so that
the final structure is self-consistent.

\section{Line list data}
\subsection{Atomic lines}

For the atomic lines, we use the database from Kurucz
\citep{2017CaJPh..95..825K,1992RMxAA..23...45K} including all updates up to 2019.
This results in a database of about 851 million lines from a total of 389 atoms and ions from H~I up to U~II
(with gaps). The largest contributor by number of lines is Rh~V with about 37.5 million lines or 4.4\% of the total number of lines,
however, this represents a negligible source of opacity for the conditions that we consider in this model grid.
For the line opacities, the opacities due to iron-group elements such
as Fe, Co, Ni are the most important. In order to dynamically account for the most important opacity sources, 
we use a line selection procedure that is an evolution of the procedure used in \cite{MDpap}: for each model and  each line in the database
its opacity in the line center is compared to the local continuous (bound-free and free-free) opacity at a prescribed number of
reference layers in the current $(T,P_{\rm g})$ structure. If the
ratio of line to continuous opacity is larger
than a given parameter (in the NewEra grid this is set to $10^{-4}$) for at least one of the reference layers, the
line will be included in the calculation, otherwise it will be ignored. In addition, a further test is made so that lines with
a  ratio of line to continuum  opacity larger than a second parameter (in the NewEra grid this is set to $1$) for at least one of the
reference layers will be considered with individual Voigt profiles, whereas for the weaker lines (line/continuum $< 1$), Gauss profiles are used.
With this procedure the total number of atomic lines varies from about $8\alog{5}$ (around $\Teff=2300\K$) to $2.5\alog{6}$
(around $\Teff=5000\K$) where about 10\% of the selected lines are
very strong lines with individual Voigt profiles.

\subsection{Molecular lines}

The molecular line database is significantly different compared to the \cite{Husser_etal_2013} model grid. 
The vast majority of the molecular lines are from the Exomol database \citep{2016JMoSp.327...73T}.
The full list of all molecular lines and their sources used in this work are given in \autoref{tab:molecules}.

\begin{table*}[ht]
 \caption[]{\label{tab:molecules}Molecular lines and their sources used in the NewEra model grid.}
\begin{tabular}{lc}
 \hline \hline
Molecules &  References \\\hline
$ ^{16}$O$_3$,
$ ^{16}$O$^{16}$O$^{18}$O,
$ ^{16}$O$^{18}$O$^{16}$O,
$ ^{16}$O$^{16}$O$^{17}$O,
$ ^{16}$O$^{17}$O$^{16}$O,
$ ^{14}$N$^{14}$N$^{16}$O,
$ ^{14}$N$^{15}$N$^{16}$O,
$ ^{15}$N$^{14}$N$^{16}$O,
$ ^{14}$N$^{14}$N$^{18}$O & 1\\
$ ^{14}$N$^{14}$N$^{17}$O,
$ ^{12}$C$^{17}$O,
$ ^{14}$N$^{16}$O$_2$,
$ ^{16}$OH,
$ ^{18}$OH,
$ ^{16}$OD,
H$^{12}$C$^{15}$N,
$ ^{12}$C$^{12}$CH$_6$ & 1 \\
$ ^{12}$C$^{16}$O$^{19}$F,
$ ^{32}$S$^{19}$F$_{6}$,
H$_2 ^{34}$S,
H$_2 ^{33}$S,
H$^{16}$O$_2$,
$ ^{12}$C$^{13}$CH$_4$ & 1\\
$ ^{12}$C$^{16}$O$_2$,  $ ^{13}$C$^{16}$O$_2$, $ ^{12}$C$^{16}$O$^{18}$O,
$ ^{12}$C$^{16}$O$^{17}$O, $ ^{12}$C$^{18}$O$_2$, $
^{12}$C$^{17}$O$^{18}$O, $ ^{13}$C$^{18}$O$_2$ & 2\\
$ ^{9}$BeH, $ ^{9}$Be$^{2}$H, $ ^{9}$Be$^{3}$H & 13\\
$ ^{24}$Mg$^{16}$O, $ ^{24}$Mg$^{17}$O, $ ^{24}$Mg$^{18}$O, $
^{25}$Mg$^{16}$O, $ ^{26}$Mg$^{16}$O & 14 \\
$ ^{26}$Al$^{16}$O, $ ^{27}$Al$^{16}$O, $ ^{27}$Al$^{17}$O, $
^{27}$Al$^{18}$O & 14\\
 $ ^{46}$Ti$^{16}$O, $ ^{47}$Ti$^{16}$O, $ ^{48}$Ti$^{16}$O,
$^{49}$Ti$^{16}$O, $ ^{50}$Ti$^{16}$O & 15\\
$ ^{6}$Li$^{35}$Cl, $ ^{6}$Li$^{37}$Cl, $ ^{7}$Li$^{35}$Cl, $
^{7}$Li$^{37}$Cl & 6,21\\
$ ^{12}$C$^{32}$S, $ ^{12}$C$^{33}$S, $ ^{12}$C$^{34}$S, $ ^{12}$C$^{36}$S,
$ ^{13}$C$^{32}$S, $ ^{13}$C$^{33}$S, $ ^{13}$C$^{34}$S, $
^{13}$C$^{36}$S & 27\\
$ ^{14}$N$^{16}$O, $ ^{14}$N$^{17}$O, $ ^{14}$N$^{18}$O,
$ ^{15}$N$^{16}$O, $ ^{15}$N$^{17}$O, $ ^{15}$N$^{18}$O & 28\\
$ ^{14}$N$^{32}$S, $ ^{14}$N$^{33}$S, $ ^{14}$N$^{34}$S, $
^{14}$N$^{36}$S, $ ^{15}$N$^{32}$S & 32\\
$ ^{28}$Si$^{32}$S, $ ^{28}$Si$^{33}$S, $ ^{28}$Si$^{34}$S,$ ^{28}$Si$^{36}$S,
$ ^{29}$Si$^{32}$S, $ ^{29}$Si$^{33}$S, $ ^{29}$Si$^{34}$S,$ ^{29}$Si$^{36}$S,
$ ^{30}$Si$^{32}$S, $ ^{30}$Si$^{33}$S, $ ^{30}$Si$^{34}$S,$
^{30}$Si$^{36}$S & 37\\
$ ^{40}$Ca$^{19}$F, $ ^{42}$Ca$^{19}$F, $ ^{43}$Ca$^{19}$F, $
^{44}$Ca$^{19}$F, $ ^{46}$Ca$^{19}$F, $ ^{48}$Ca$^{19}$F & 6,44\\
$ ^{39}$K$^{35}$Cl, $ ^{39}$K$^{37}$Cl,
$ ^{41}$K$^{35}$Cl, $ ^{41}$K$^{37}$Cl& 45 \\
$ ^{24}$Mg$^{19}$F, $ ^{25}$Mg$^{19}$F, $ ^{26}$Mg$^{19}$F & 6,35 \\
$ ^{28}$Si$^{16}$O, $ ^{28}$Si$^{17}$O, $ ^{28}$Si$^{18}$O, $
^{29}$Si$^{16}$O, $ ^{30}$Si$^{16}$O & 57\\
D$^{35}$Cl, D$^{37}$Cl, H$^{35}$Cl,  H$^{37}$Cl & 18\\
$ ^{12}$C$^{14}$N, $ ^{12}$C$^{15}$N, $ ^{13}$C$^{14}$N, $
^{13}$C$^{15}$N & 72,73\\
$ ^{12}$C$^{16}$O, $ ^{12}$C$^{18}$O, $ ^{13}$C$^{16}$O, $
^{13}$C$^{17}$O, $ ^{13}$C$^{18}$O & 74\\
$ ^{28}$Si$^{14}$N, $ ^{28}$Si$^{15}$N, $ ^{29}$Si$^{14}$N, $
^{30}$Si$^{14}$N & 84\\
$ ^{32}$SH, $ ^{32}$SD, $ ^{33}$SH, $ ^{34}$SH, $ ^{36}$SH & 88\\
\hline
\hline
\end{tabular}
\end{table*}

\begin{table*}[ht]
 \caption[]{\label{tab:molecules2}Molecules used in this work (cont'd).}
\begin{tabular}{lclclc}
 \hline \hline
Molecules &  References &Molecules &  References &Molecules &
References  \\\hline
CH$_4$ & 3,4 & $ ^{51}$V$^{16}$O & 5 & $ ^{56}$FeH & 6,7,8 \\
$^{23}$NaH, $ ^{23}$NaD & 9 &
$ ^{26}$AlH, $ ^{27}$AlH, $ ^{27}$AlD & 10  &
$ ^{45}$ScH & 11\\
 $^{7}$LiH & 12 &
$ ^{48}$TiH &  6,7 &
$ ^{52}$CrH & 6 \\
H$^{12}$C$^{14}$N & 16 &
H$^{13}$C$^{14}$N & 17 &
H$^{35}$Cl & 18 \\
H$_2^{32}$S & 19,20 &
$^{6}$Li$^{19}$F, $ ^{7}$Li$^{19}$F & 6,21 &
$ ^{12}$C$_2$, $ ^{12}$C$^{13}$C, $ ^{13}$C$_2$ & 22,23\\
$ ^{12}$CH, $ ^{13}$CH & 6,24 &
$ ^{12}$C$^{31}$P & 6,25,26 &
$ ^{14}$NH & 6,29,30,31\\
$ ^{23}$Na$^{19}$F & 33 &
$ ^{23}$Na$^{35}$Cl, $ ^{23}$Na$^{37}$Cl & 34 &
$ ^{24}$Mg$^{19}$F, $ ^{26}$Mg$^{19}$F & 6,35\\
$ ^{27}$Al$^{19}$F & 6 &
$ ^{27}$Al$^{35}$Cl, $ ^{27}$Al$^{37}$Cl & 6 &
$ ^{28}$SiH, $ ^{28}$SiD, $ ^{29}$SiH, $ ^{30}$SiH & 36\\
$ ^{31}$P$^{14}$N, $ ^{31}$P$^{15}$N & 38 &
$ ^{31}$P$^{16}$O & 39 & 
$ ^{31}$PH & 40\\
$ ^{31}$P$^{32}$S & 41 & 
$ ^{32}$S$^{16}$O$_2$ & 42 & 
$ ^{32}$SH, $ ^{32}$SD,$ ^{33}$SH, $ ^{34}$SH, $ ^{36}$SH & 42\\
$ ^{39}$K$^{19}$F, $ ^{41}$K$^{19}$F & 6,33 &
H$_2^{16}$O & 46 &
H$_2^{16}$O$_2$ & 47\\
$ ^{15}$NH$_3$ & 48 &
$ ^{14}$NH$_3$ & 49,50 &
H$^{14}$N$^{16}$O$_3$ & 51\\
H$_2^{12}$C$^{16}$O & 52 &
$ ^{12}$C$_2$H$_2$  & 53 & 
$ ^{12}$C$_2$H$_4$ & 54\\
$ ^{12}$CH$_3$F & 55 &
$ ^{12}$CH$_3^{35}$Cl, $ ^{12}$CH$_3^{37}$Cl & 56 &
$ ^{28}$SiH$_4$ & 58 \\
$ ^{32}$S$^{16}$O$_3$ & 59 &
$ ^{31}$PH$_3$ & 60 &
$ ^{12}$C$^{16}$O$_2$ & 61\\
$ ^{40}$CaH & 62 &
$ ^{24}$MgH, $ ^{25}$MgH, $ ^{26}$MgH & 62 &
H$_2$ & 64 \\
HD & 64 &
H$_2^{17}$O,H$_2^{18}$O & 65 &
HD$^{16}$O & 66 \\
H$^{19}$F & 67,68,69 &
$ ^{12}$CH$_3$& 70 &
$ ^{12}$CH$_4$& 4,71 \\
$ ^{14}$NH & 30,6,75,76 &
$ ^{14}$N$_2$ & 77,78,79 &
$ ^{14}$N$^{16}$O & 80\\
$ ^{14}$Na$^{16}$OH  & 81 &
$ ^{14}$Na$^{16}$O & 82 &
$ ^{28}$SiH$_2$ & 83 \\
$ ^{28}$Si$^{16}$O & 85 &
$ ^{28}$Si$^{16}$O$_2$ & 86 &
$ ^{31}$P$^{19}$F$_3$ & 87 \\
$ ^{39}$K$^{16}$OH & 81 &
$ ^{40}$Ca$^{16}$O & 90 &
$ ^{40}$Ca$^{16}$OH & 91\\
$ ^{56}$FeH & 6,8 &
$ ^{75}$AsH$_3$ & 91 &
$ ^{89}$Y$^{16}$O & 92\\
$ ^{139}$Y$^{16}$O & 93 &
HD$^+$ & 12,62 &
H$_2$D$^+$ & 94\\
H$_3^+$& 95 &
H$_3^{16}$O$^+$ & 96 &
$ ^3$HeH$^+$, $ ^3$HeD$^+$, $ ^4$HeH$^+$, $ ^4$HeD$^+$ & 64 \\
$ ^7$LeH$^+$ & 12 &
$ ^{16}$OH$^+$ & 6,97 &
$ ^{90-96}$ZrO & 98\\
\hline
\hline
\end{tabular}
\tablebib{
(1) \citet{2009JQSRT.110..533R}
(2) \citet{2010JQSRT.111.2139R}.
(3) \citet{YURCHENKO201369} (4)  \cite{ExomolIV2014};
(5) \citet{ExomolXVIII2016};
(6) \citet{BERNATH2020106687}; (7) \cite{2005ApJ...624..988B}; (8) \cite{Dulick_2003};
(9) \citet{ExomolX2015};
(10) \citet{Exomol_XXVIII_2018};
(11) \citet{Lodi_etal_2015};
(12) \citet{Coppola_etal_2011};
(13) \citet{2018JPhB...51r5701D};
(14) \citet{Exomol_XXXII_2019};
(15) \citet{Exomol_XXXIII_2019}
(16) \citet{Harris_etal_2006};
(17) \citet{Exomol_III_2014};
(18) \citet{2017JQSRT.203....3G};
(19) \cite{Exomol_XVI_2016}; (20) \cite{2018JQSRT.218..178C};
(21) \cite{Bittner_2018};
(22) \citet{Exomol_XXXI_2018}; (23) \cite{2020MNRAS.497.1081M};
(24) \citet{Masseron_2014};
(25) \citet{RAM2014107}; (26) \cite{QIN2021107352};
(27) \citet{Exomol_XII_2015};
(28) \citet{ExoMol_XXI_2017};
(29) \cite{2014JChPh.141e4310B};
(30) \cite{FERNANDO201829}; (31) \cite{2015JChPh.143b6101B};
(32) \citet{ExoMol_XXVI_2018};
(33) \citet{FROHMAN2016104};
(34) \citet{Exomol_V_2014};
(35)  \citet{HOU2017511};
(36) \citet{Exomol_XXIV_2018};
(37) \citet{Exomol_XXV_2018};
(38) \citet{Exomol_VI_2014};
(39) \citet{Exomol_XXIII_2017};
(40) \citet{Exomol_XXXIV_2019};
(41) \citet{Exomol_XXIII_2017};
(42) \citet{Exomol_XIV_2016};
(43) \citet{ExoMol_XXVI_2018};
(44) \cite{HOU201844};
(45) \citet{Exomol_V_2014};
(46) \citet{Exomol_XXX_2018};
(47) \citet{Exomol_XV_2016};
(48) \citet{YURCHENKO201528};
(49) \citet{ALDERZI2015117}; (50) \cite{Exomol_XXXV_2019};
(51) \citet{Exomol_XI_2015};
(52) \citet{ExoMol_VIII_2015};
(53) \citet{ExoMol_XXXVII_2020};
(54) \citet{ExoMol_XXVII_2018};
(55) \citet{ExoMol_XXVII_2018};
(56) \citet{ExoMol_XXIX_2018};
(57) \citet{ExoMol_II_2013};
(58) \citet{ExoMol_XXII_2017};
(59) \citet{ExoMol_XVII_2016};
(60) \citet{ExoMol_VII_2015};
(61) \citet{ExoMol_XXXIX_2020};
(62) \citet{ExoMol_XLV_2022};
(63) \citet{2019A&A...630A..58R};
(64) \citet{Amaral_2019};
(65) \citet{Exomol_XXX_2018};
(66) \citet{Voronin_etal_2010};
(67) \citet{LI201378}; (68)  \cite{COXON2015133}; (69) \cite{2021JChPh.155u4303S};
(70) \citet{2019JPCA..123.4755A};
(71) \cite{2017A&A...605A..95Y};
(72) \citet{Brooke_2014}; (73) \cite{2021MNRAS.505.4383S};
(74) \citet{Li_2015};
(75) \cite{2014JChPh.141e4310B}; (76)  \cite{2015JChPh.143b6101B};
(77) \citet{WESTERN2018127}; (78) \cite{WESTERN2017221}; (79)  \cite{1969JChPh..51..689S};
(80) \citet{ExoMol_XLII_2021};
(81) \citet{ExoMol_XLI_2021};
(82) \citet{ExoMol_XLIII_2022};
(83) \cite{CLARK2020106929};
(84) \citet{2022MNRAS.516.1158S};
(85) \citet{ExoMol_XLIV_2022};
(86) \citet{ExoMol_XXXVIII_2020};
(87) \citet{Mant_etal_2020};
(88) \citet{ExoMol_XXXVI_2019};
(89) \citet{ExoMol_XIII_2016};
(90) \citet{2022MNRAS.516.3995O};
(91) \citet{2019PCCP...21.3264C};
(92) \citet{2019PCCP...2122794S};
(93) \citet{Bernath_2022};
(94) \citet{Sochi_2010};
(95) \citet{ExoMol_XX_2017}; 
(96) \citet{ExoMol_XL_2020};
(97) \cite{2017ApJ...840...81H};
(98) \cite{zrolines}.
}
\end{table*}

\begin{figure}[h]
    \centering
    \includegraphics[width=\columnwidth]{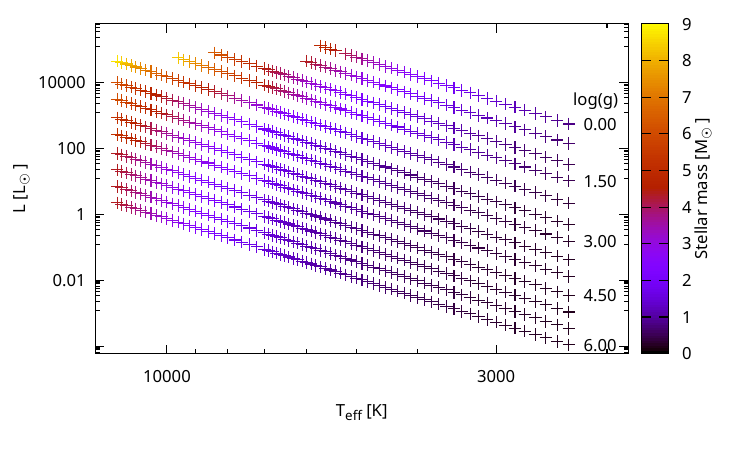}
    \caption{Available models for solar metallicity in a Hertzsprung-Russell Diagram.
    The parameters of the models are the effective temperature (on the x-axis), $\logg$ (indicated next to the series of models for constant $\logg$), the metallicity (fixed to [M/H]=0 in this plot) and stellar mass (color coded) following \citet{Husser_etal_2013}. The luminosity of the models follows from the given parameters.
    See \autoref{sec:modelgrid} for a discussion of the shape formed by the available models.}
    \label{fig:LTeffgrid}
\end{figure}

\begin{figure}[h]
    \centering
    \includegraphics[width=\columnwidth]{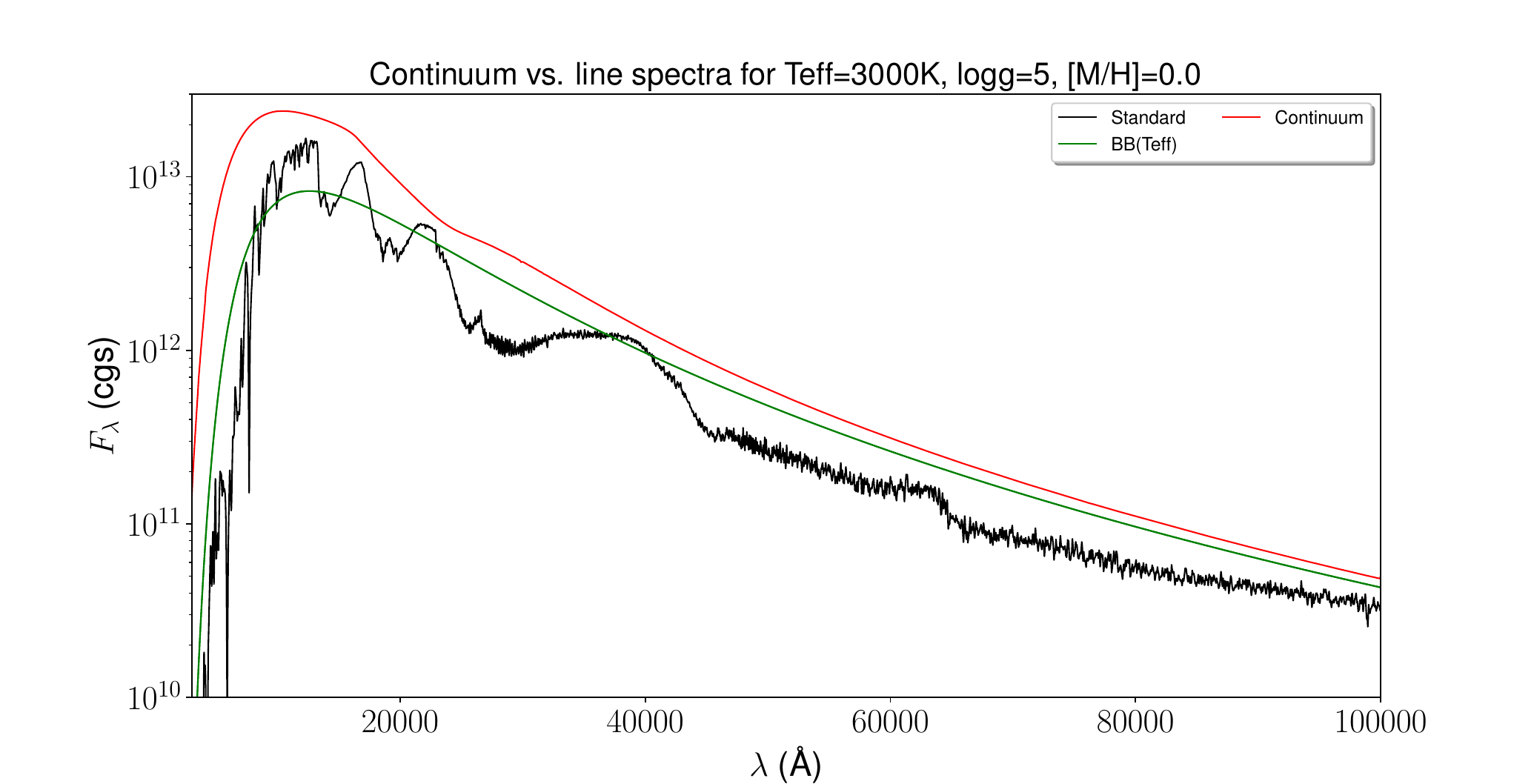}
    \caption{\AAmarker{Effects of line blanketing on the spectral energy distribution for a NewEra
    grid model with $\Teff=3000$\,\K, $\logg=5.0$ and solar abundances (black line). The structure of the grid model 
    was \AAmarker{used} to compute the 'true continuum', i.e., the SED without any spectral lines included (red line). 
    In addition, the blackbody for the effective temperature is shown in the green curve.}}
    \label{fig:Continuum}
\end{figure}

The combined molecular line list contains about 823.8 billion individual lines with transition
probabilities $>0$. 
In order to reduce the size
of the line list and to save processing time, we have created a sub-list by the following procedure:
for each of the Exomol species and for each provided wavelength band data file individually, we estimate the total
line opacity at infinite temperature per molecule and include all lines that contribute more than $10^{-3}$ to $10^{-4}$
(depending on species) to the estimated total. This procedure results in a list with about 228.9 billion lines. 
Furthermore, for the model parameters covered by the grid presented here, in particular for $\Teff\ge 2300\,$K, 
a number of molecules are not present in the atmospheres. When we omit these lines from the list, we obtain
a small list with only about 20.5 billion lines.  At a nominal resolution of $25\ang$ the differences between spectra of a model ($\Teff=2300\K$, $\logg=4.5$, solar abundances) with the full (823.8G lines) and the smallest (20.5G) line lists peak around $4\mu$m with a maximum flux difference less than $0.4\%$.

\section{Model grid description}

\label{sec:modelgrid}
The new model grid follows the parameters and setup of \citet{Husser_etal_2013}, this 
includes the $\Teff$--$\logg$ -- (mixing length, mass) relations\AAmarker{, the base solar abundances used,} and the abundance pattern 
variations. 
The overall parameter range provided is $2300\K \le \Teff \le 12000\K$ (with $100\K$ steps below $8000\K$ and $200\K$ steps
above $8000\K$), $0.0\le \logg \le 6.0$ (with steps of $0.5$ dex) and metallicities [M/H] from $-4.0$ to $+0.5$ in steps of $0.5$ dex. 
For [M/H] from $-2.0$ to $0.0$ we provide additional $\alpha$ element variations from $-0.2$ to $+1.2$ with $0.2$ dex steps for a subset of the $(\Teff, \logg)$ parameters of the model grid. 

The model grid is not `square' in the sense that for any possible parameter combination there is a
model available. For example, low gravity models will become unstable against radiation pressure depending
on effective temperature and composition. Such non-static models are not included in this grid. 
For the solar metallicity ([M/H]=0.0) subset, \autoref{fig:LTeffgrid} shows which models are available.
Overall, \nmodel models are included with
the data described below. 

\subsection{Code checks}

Computing a model grid on these scales is impractical on local computers and
multiple HPC resources were used to compute models and synthetic spectra: Hummel2 (Hamburg),
HLRN-4 (Berlin and G\"ottingen), as well as NERSC (Perlmutter CPU and GPU partitions).
Gaps in the model grid (e.g., due to random hardware or I/O problems on the
supercomputers) were filled using local resources (M1 Mac Studio, M1 Mac Mini,
Linux Xeon PC).  Thus, it is very important to verify that these very different
systems all produce  the same model structure.  We have run a number of tests
on all  systems in order to verify that the results for model iterations agree
to better than $10^{-6}$ relative accuracy in all quantities saved in the model
restart file (e.g., temperatures, gas \& electron pressures, densities, opacity
averages etc.).  These tests included serial and MPI runs (with up to 512 MPI
processes), different compilers: gfortran, Intel ifort and ifx, NVIDIA nfortran
CPU and GPU (OpenACC and OpenMP/offload), AMD flang and NEC fortran/VE on
different CPU/GPU architectures: Apple M1, X86\_64, AMD64, NVIDIA A100 GPU, NEC
VE.  During these tests a number of improvements were implemented (e.g., to
make the results of the line selection process completely consistent for
different MPI setups) before the model grid was computed.

\begin{figure*}[h]
\begin{minipage}{0.5\textwidth}
\centering
  \includegraphics[width=\textwidth]{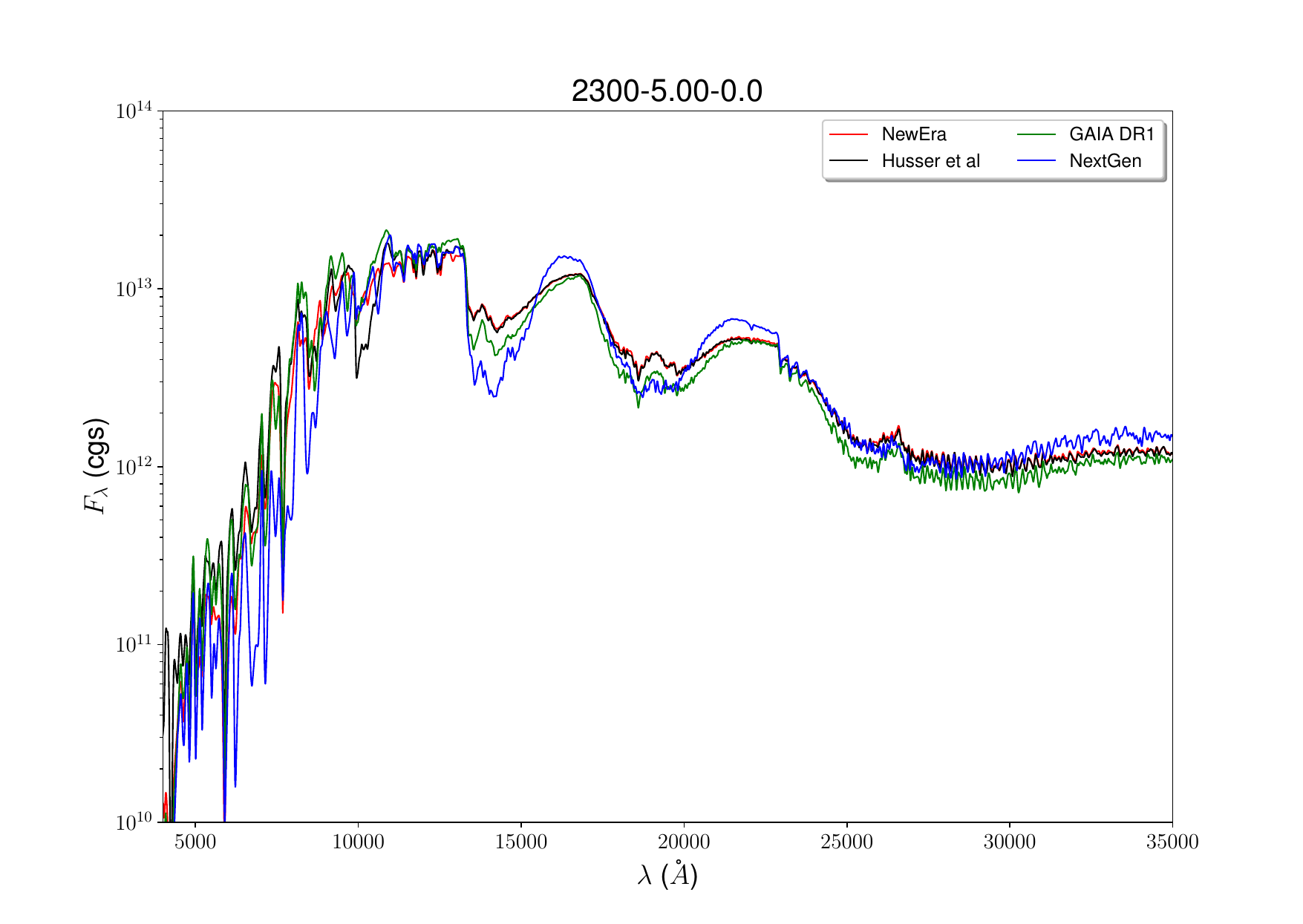}
\end{minipage}
\begin{minipage}{0.5\textwidth}
\centering
  \includegraphics[width=\textwidth]{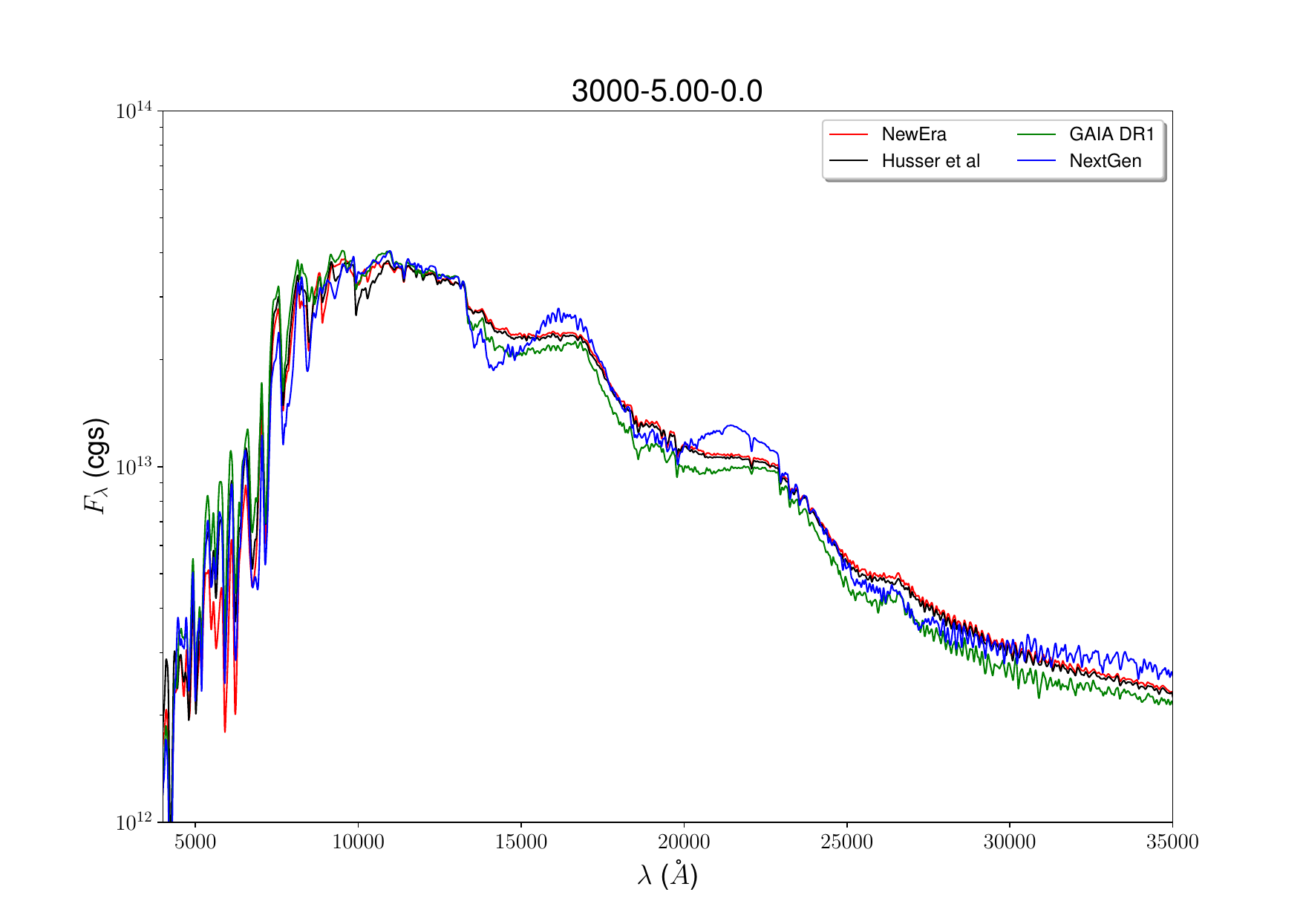}
\end{minipage}\\
\begin{minipage}{0.5\textwidth}
\centering
  \includegraphics[width=\textwidth]{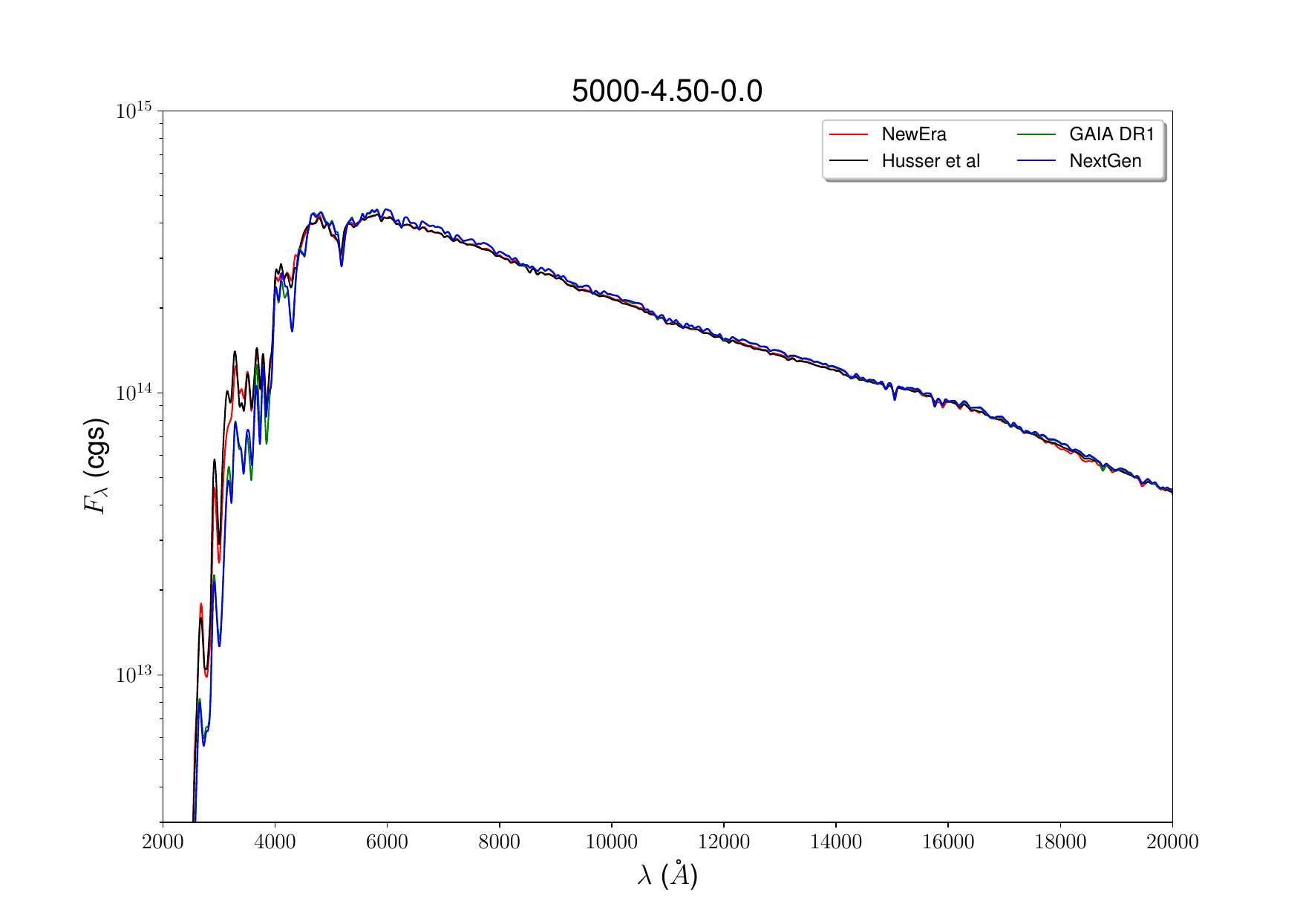}
\end{minipage}
\begin{minipage}{0.5\textwidth}
\centering
  \includegraphics[width=\textwidth]{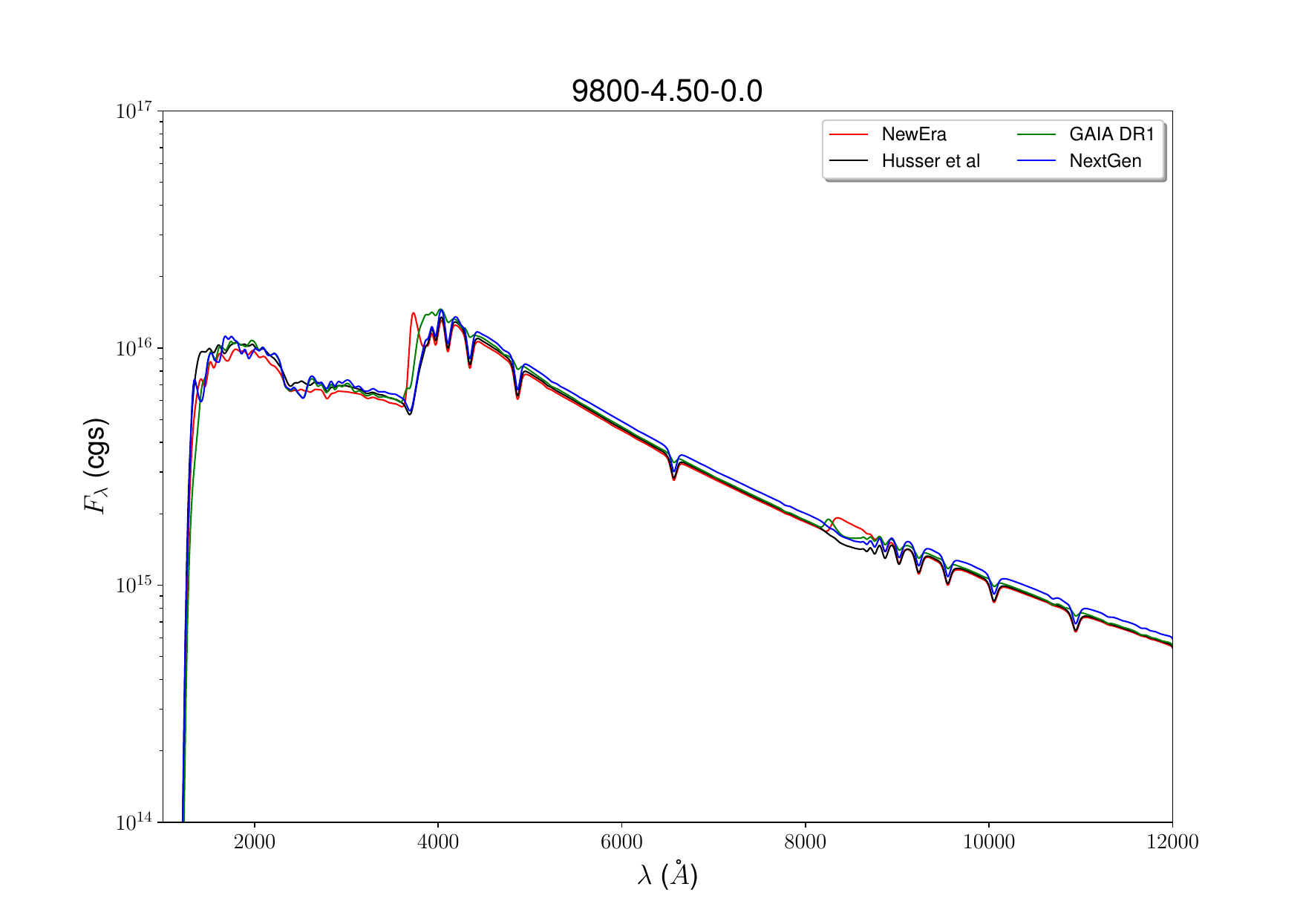}
\end{minipage}

\caption{The grid spectra for model with
  \AAmarker{$\Teff = 2300, 3000, 5000, 9800\,\K$  and $\logg = 5.0,\ 5.0,\ 4.5,$ and
  $4.5$} for \AAmarker{solar metallicity} respectively, compared \AAmarker{to} the NextGen \citep{ng-hot},
  GAIA DR1 \citep{2005A&A...442..281K}, ACES \citep{Husser_etal_2013}, and New Era
  (this work). 
  } 
\label{fig:spectra plots}
\end{figure*}

\begin{figure*}[h]
\begin{minipage}{0.5\textwidth}
\centering
  \includegraphics[width=\textwidth]{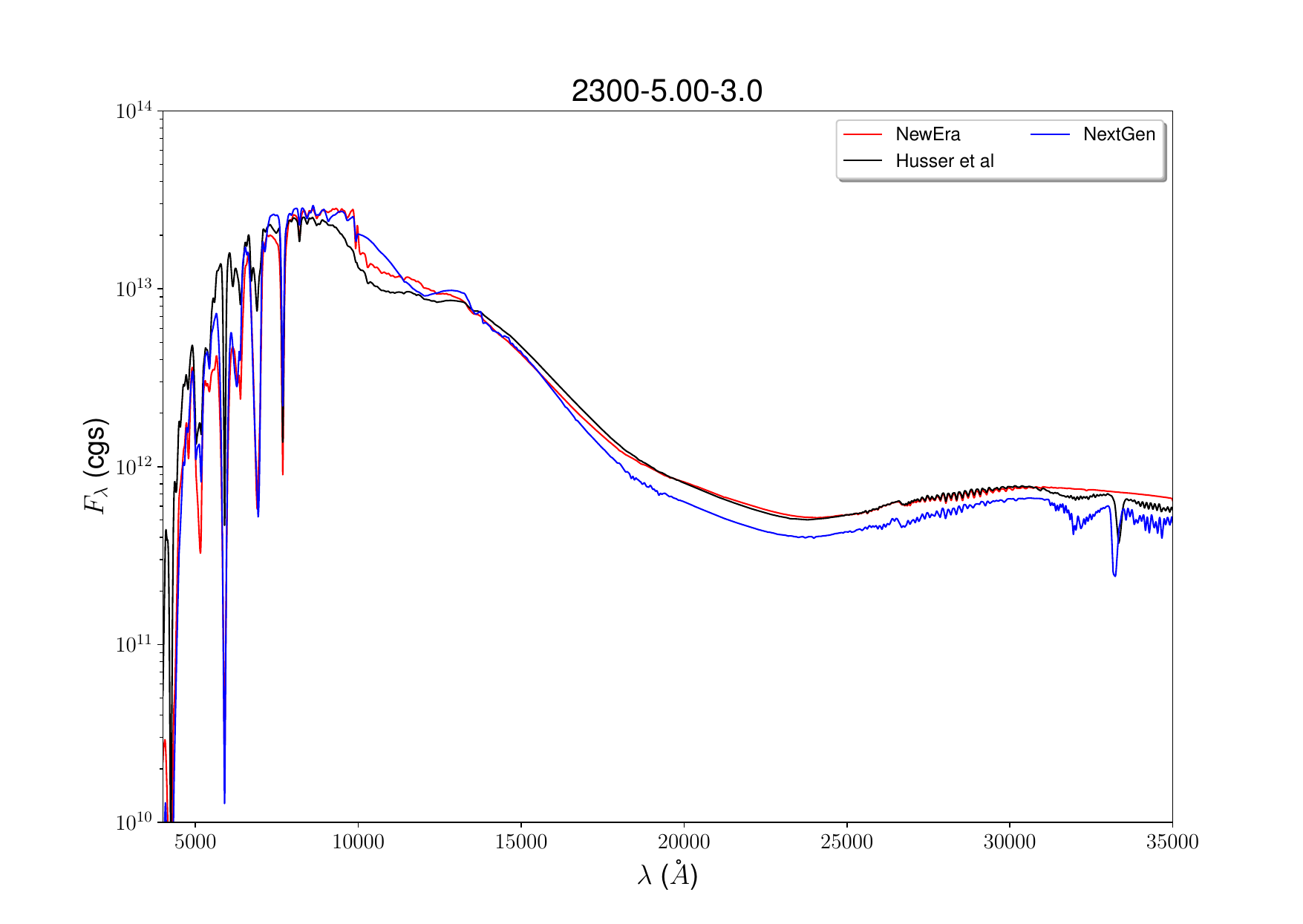}
\end{minipage}
\begin{minipage}{0.5\textwidth}
\centering
  \includegraphics[width=\textwidth]{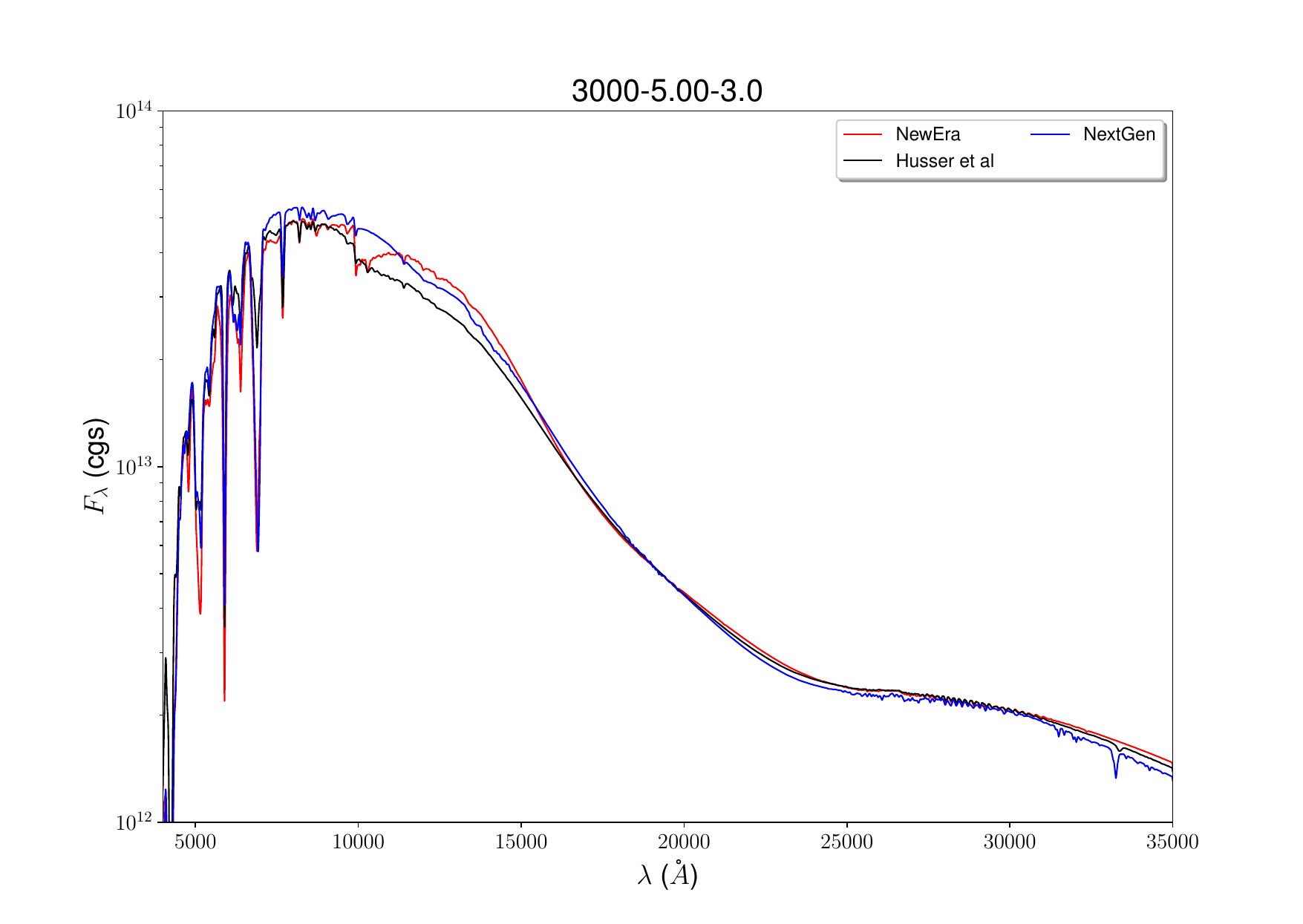}
\end{minipage}\\
\begin{minipage}{0.5\textwidth}
\centering
  \includegraphics[width=\textwidth]{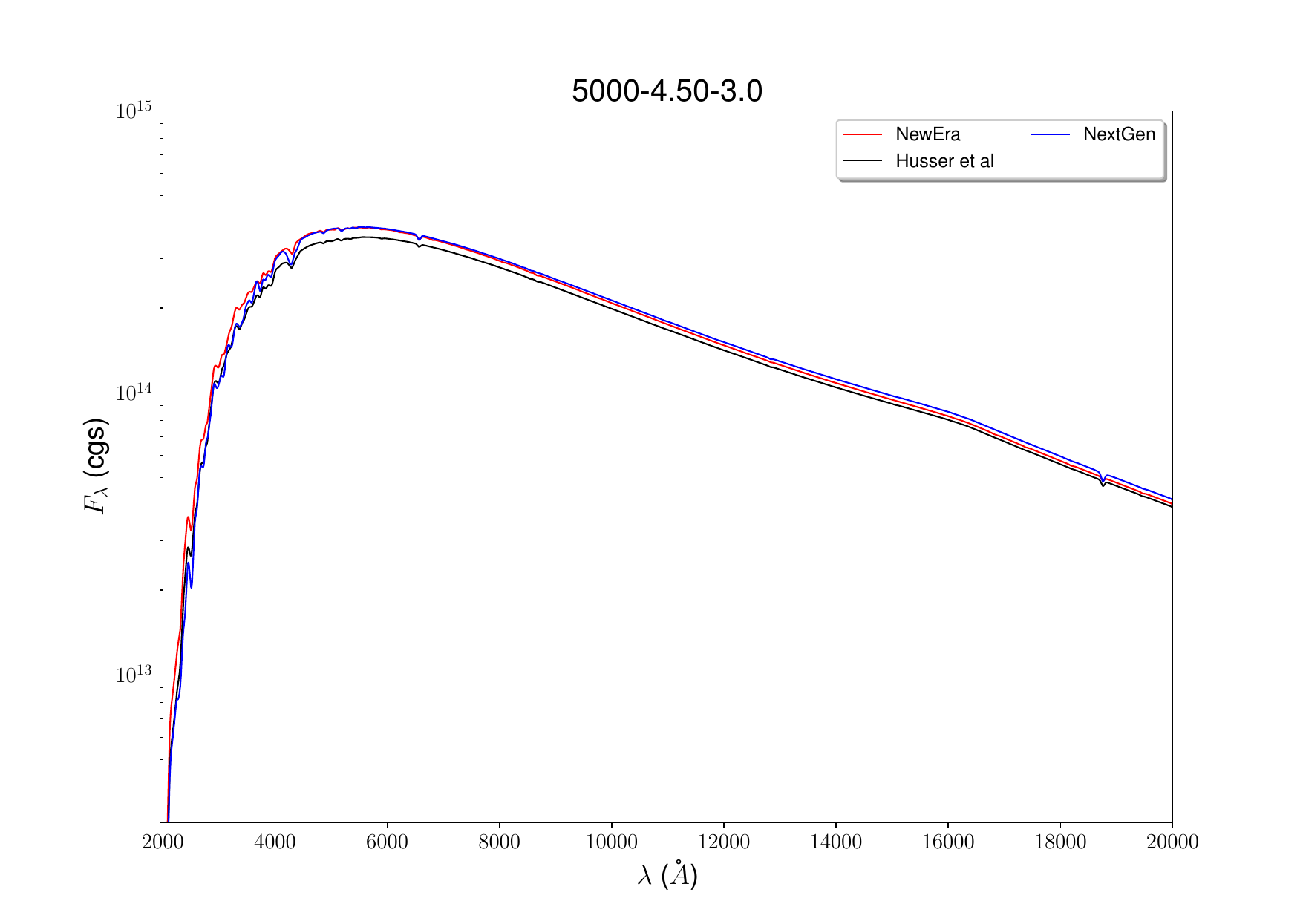}
\end{minipage}
\begin{minipage}{0.5\textwidth}
\centering
  \includegraphics[width=\textwidth]{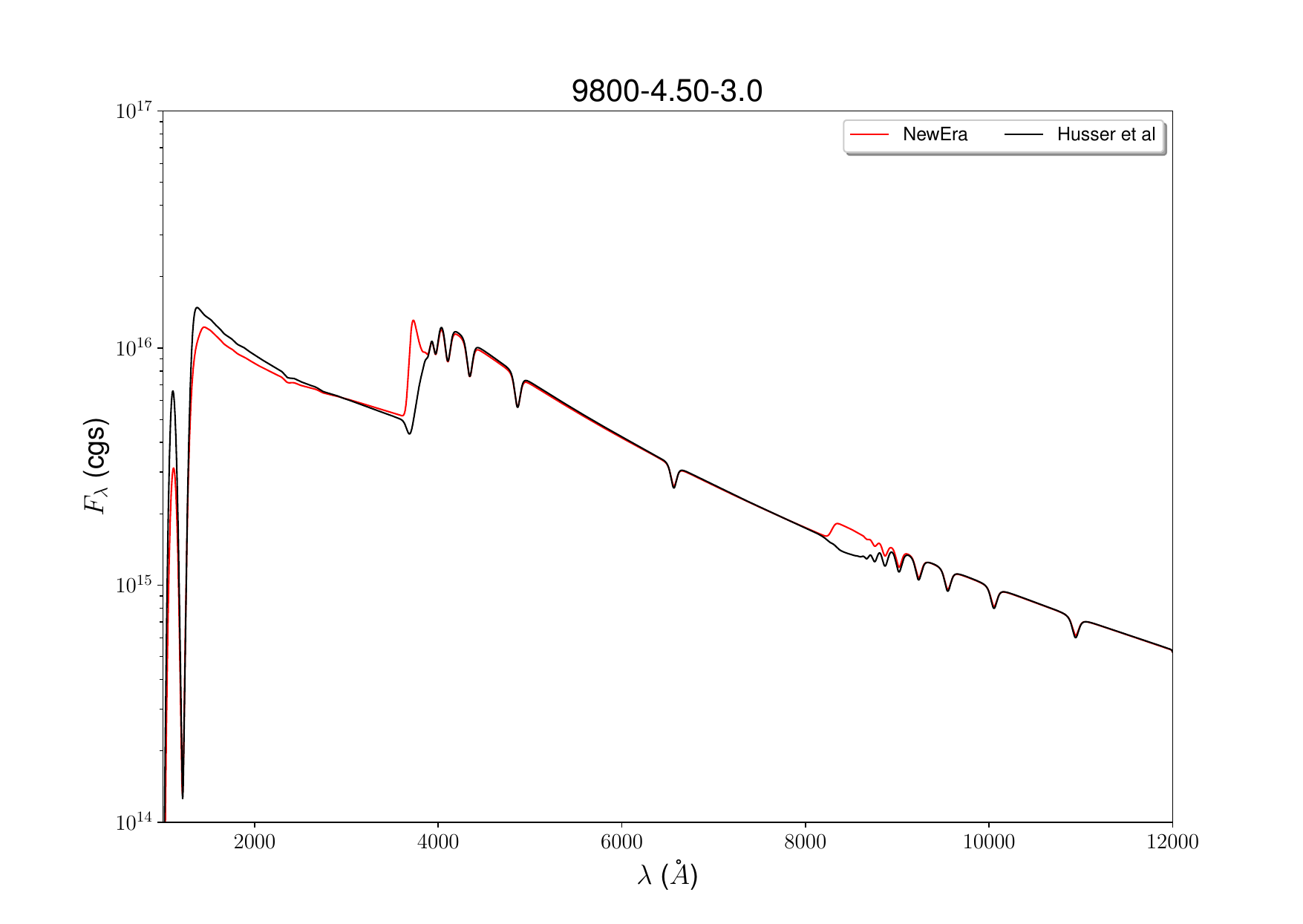}
\end{minipage}
\caption{\AAmarker{The grid spectra for models with
  \AAmarker{$\Teff = 2300, 3000, 5000, 9800\,\K$  and $\logg = 5.0,\ 5.0,\ 4.5,$ and
  $4.5$} for \AAmarker{[M/H]}=-3.0 respectively, compared to the NextGen \citep{ng-hot},
  GAIA DR1 \citep{2005A&A...442..281K}, ACES \citep{Husser_etal_2013}, and New Era
  (this work). }
  } 
\label{fig:spectra plots-3.0}
\end{figure*}

\begin{figure}[h]
\centering
  \includegraphics[width=\columnwidth]{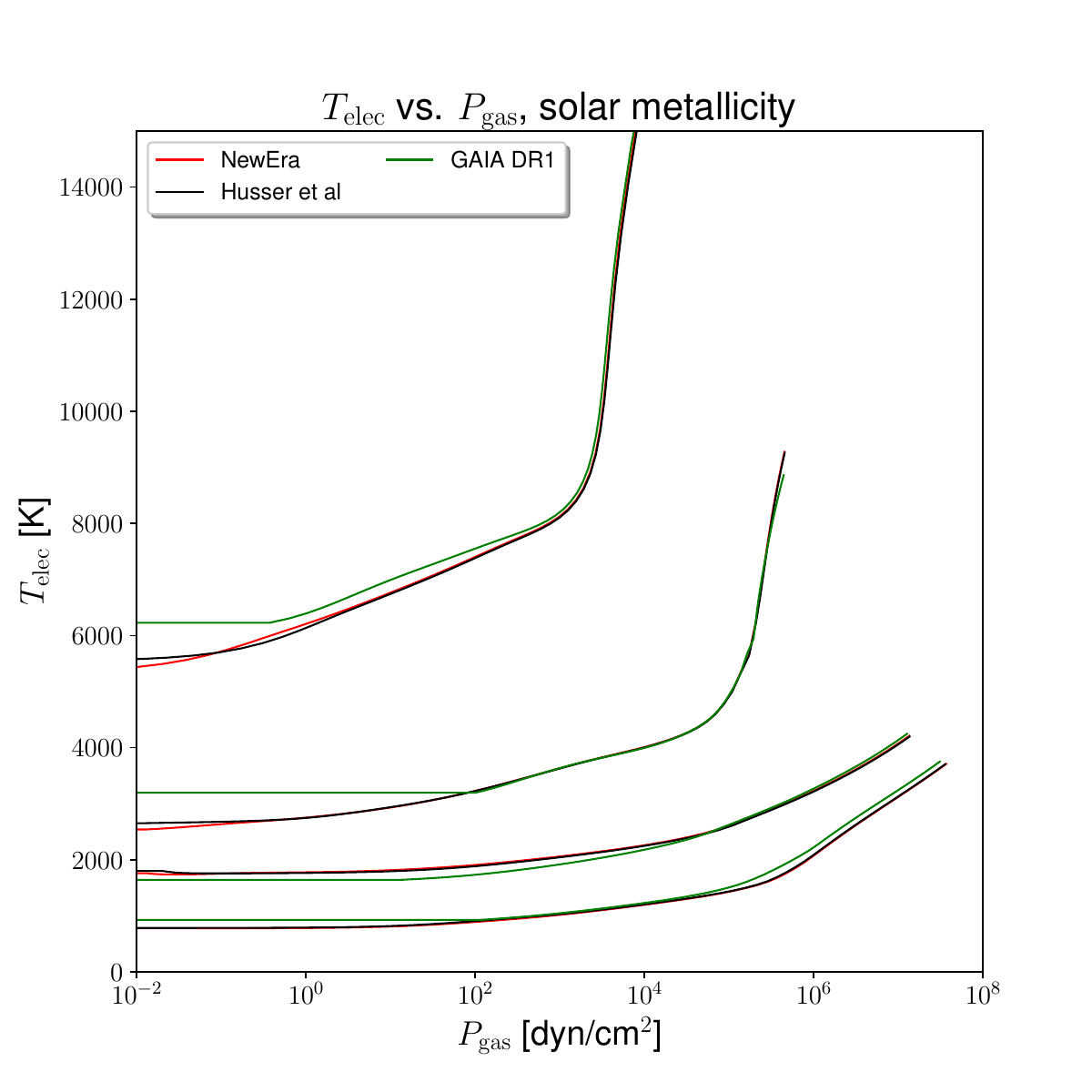}
\caption{\AAmarker{The structure of  models with (bottom to top)
  \AAmarker{$\Teff = 2300, 3000, 5000, 9800\,\K$  and $\logg = 5.0,\ 5.0,\ 4.5,$ and
  $4.5$} for solar metallicity respectively, compared to the NextGen \citep{ng-hot},
  GAIA DR1 \citep{2005A&A...442..281K}, ACES \citep{Husser_etal_2013}, and New Era
  (this work). (Not all structures are available at all metallicities) }
  } 
\label{fig:structures1}
\end{figure}

\begin{figure}[h]
\centering
  \includegraphics[width=\columnwidth]{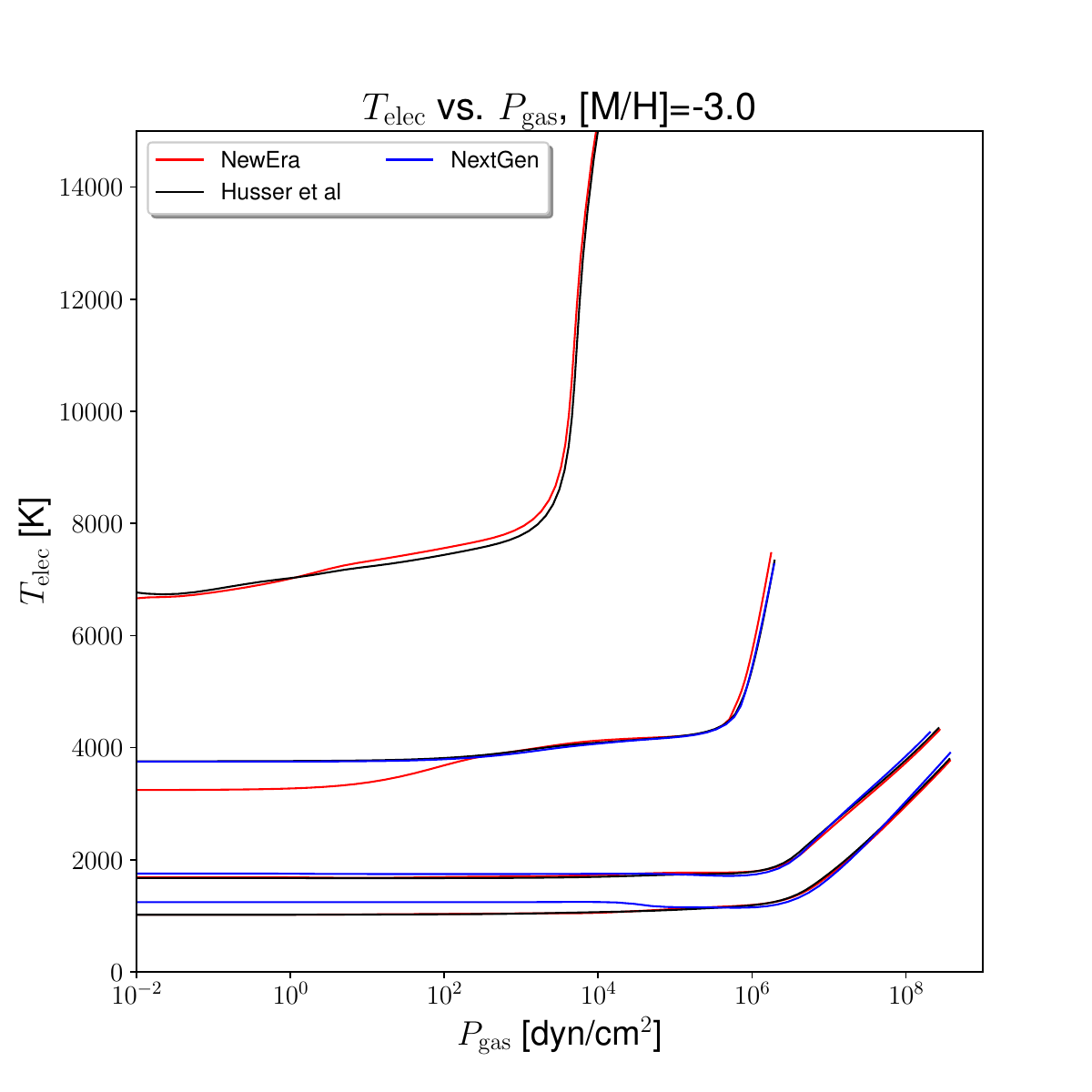}
\caption{\AAmarker{The structure of  models with (bottom to top)
  \AAmarker{$\Teff = 2300, 3000, 5000, 9800\,\K$  and $\logg = 5.0,\ 5.0,\ 4.5,$ and
  $4.5$} for \AAmarker{[M/H]}=-3.0 respectively, compared to the NextGen \citep{ng-hot},
  GAIA DR1 \citep{2005A&A...442..281K}, ACES \citep{Husser_etal_2013}, and New Era
  (this work). (Not all structures are available at all metallicities) }
  } 
\label{fig:structures2}
\end{figure}

\begin{figure}[ht]
\centering
  \includegraphics[width=\columnwidth]{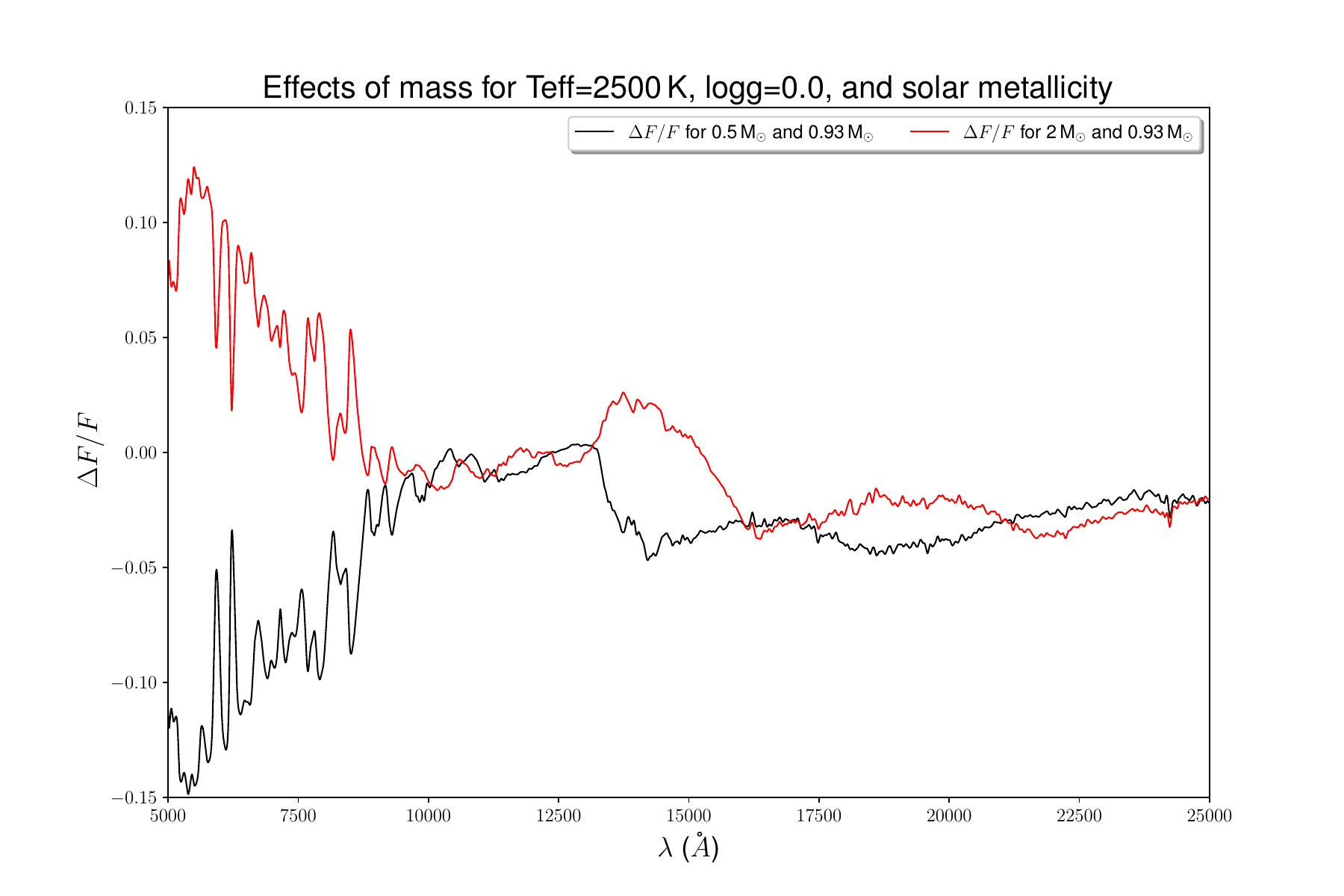}

\caption{\AAmarker{The effect of (model) mass on the spectral energy distributions (SED)
at a resolution of $25\,$\AA\ for an example model with $\Teff=2500\,$K, $\logg=0.0$, $M=0.93\,\Msun$,
and solar abundances. 
$\Delta F/F$ is defined as ($F$(mass=[0.5,2.0])-$F$(mass=0.93))/$F$(mass=0.93).}
  } 
\label{fig:mass-effects}
\end{figure}

\subsection{\AAmarker{Effects of line blanketing on the spectra}}
\AAmarker{The effects of the line blanketing on the spectral energy distribution is very large, in particular
for models with low effective temperatures. 
This is shown in \Autoref{fig:Continuum} for a model with $\Teff=3000$\,\K, clearly showing the line
blanketing effects for a large wavelength range. The 'true continuum' can differ by large factors from 
the SED with spectral lines included.}

\subsection{\AAmarker{Effects of stellar mass}}

\AAmarker{In the NewEra grid, each model is available for
one stellar mass. The effects of varying stellar mass for 
fixed ($\Teff$, $\logg$) are for example discussed in 
\cite{2017PASP..129b4201L} and \cite{2022A&A...662A..38N}. In \Autoref{fig:mass-effects}
we show the effects on the spectral energy distribution of varying the model mass by a factor 
of 4 for a low gravity NewEra grid model. The radial extension $(R_{\rm out}-R_{\rm in})/R_{\rm in} $
varies from 20\% for $0.5\Msun$ to 16\% for $0.93\Msun$ and to 10\% for $2.0\Msun$. Depending on 
the required relative accuracy, models with individualized masses may be needed for low gravities.}

\subsection{Comparison to previous grids}

In \Autoref{fig:spectra plots,fig:spectra plots-3.0} we compare representative spectra in 
the optical to near-IR for four effective temperatures to \AAmarker{up to} three previous model
generations. All spectra were reduced to a resolution of $25\,$\ang\ by convolution 
with a Gaussian for clarity. 
For $\Teff=9800\,\K$ and $5000\,\K$ the differences between the 
\citet{Husser_etal_2013} and NewEra models are small, mostly due to a different
treatment of line dissolution close to ionization edges (\AAmarker{hydrogen line dissolution at lower electron
pressures calibrated using the results of NLTE modeling of the spectrum of
Sirius A, Aufdenberg et al, (in preparation)}). At the two lower effective temperatures the differences are larger.
Here the differences between NewEra,  GAIA DR1 \citep{2005A&A...442..281K} \& NextGen  \citep{ng-hot} are due to vastly different
molecular line data, this is visible at all plotted wavelengths. In the near-IR, above
about $1.1\,\mu$m, at this low resolution the spectra of NewEra and \citet{Husser_etal_2013}
differ little. This is due to very similar water line data so that the changes at the
low resolution are not easily visible. At optical wavelengths the differences are much larger,
here the significantly different molecular line data have a large impact on the synthetic 
spectra. \AAmarker{In \Autoref{fig:structures1,fig:structures2} we plot the temperature--pressure structures for the 
different model generations. The differences between the NewEra and ACES structures
are small, indicating that the actual flux averaged opacities are quite similar whereas the
detailed spectra show significant differences.}

\subsection{Center-to-limb variation}

\begin{figure*}[h]
\begin{minipage}{0.5\textwidth}
\centering
\includegraphics[width=\textwidth]{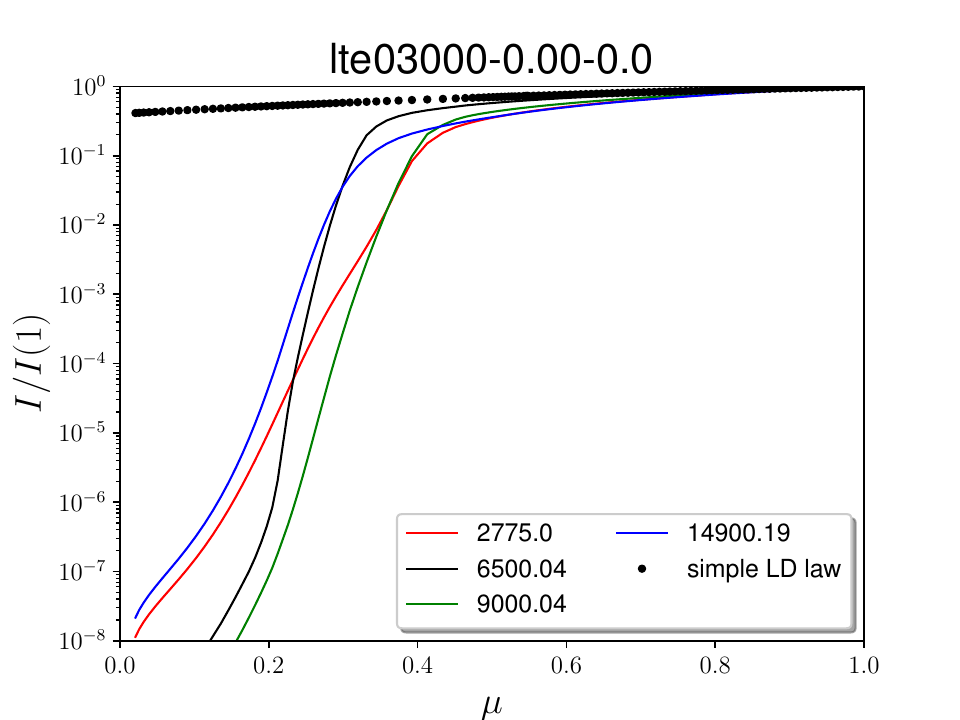}
\end{minipage}
\begin{minipage}{0.5\textwidth}
\centering
  \includegraphics[width=\textwidth]{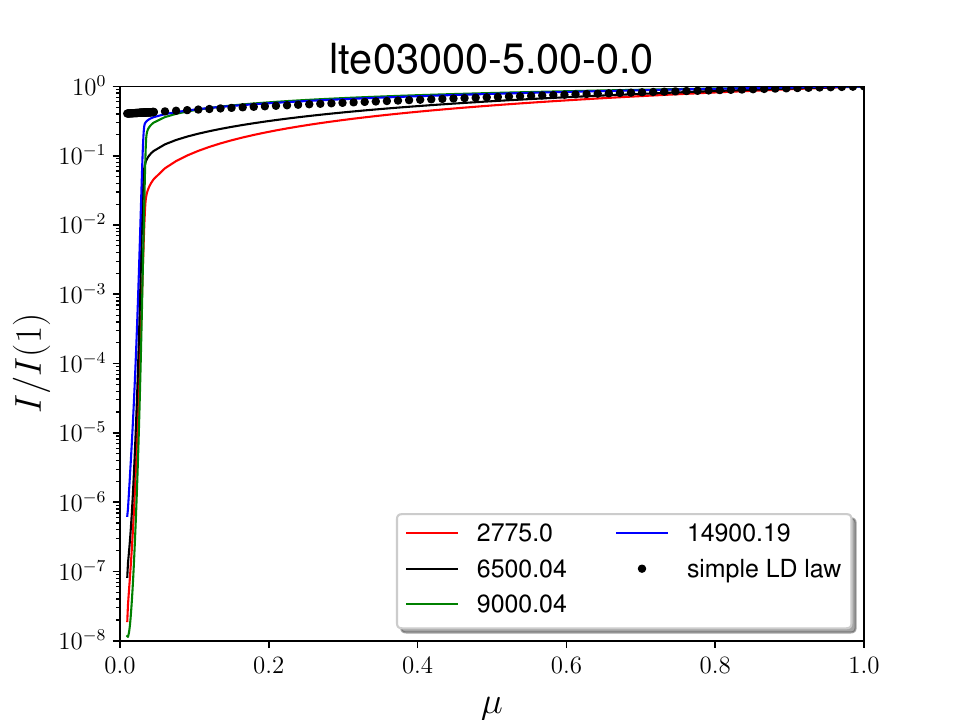}
\end{minipage}
\begin{minipage}{0.5\textwidth}
\centering
\includegraphics[width=\textwidth]{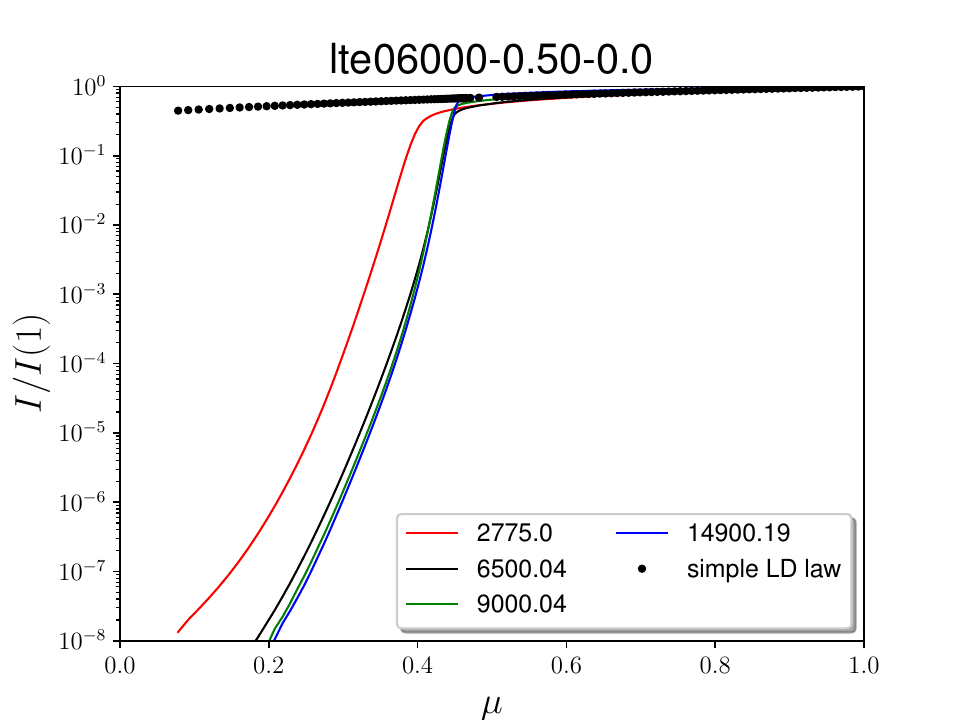}
\end{minipage}
\begin{minipage}{0.5\textwidth}
\centering
  \includegraphics[width=\textwidth]{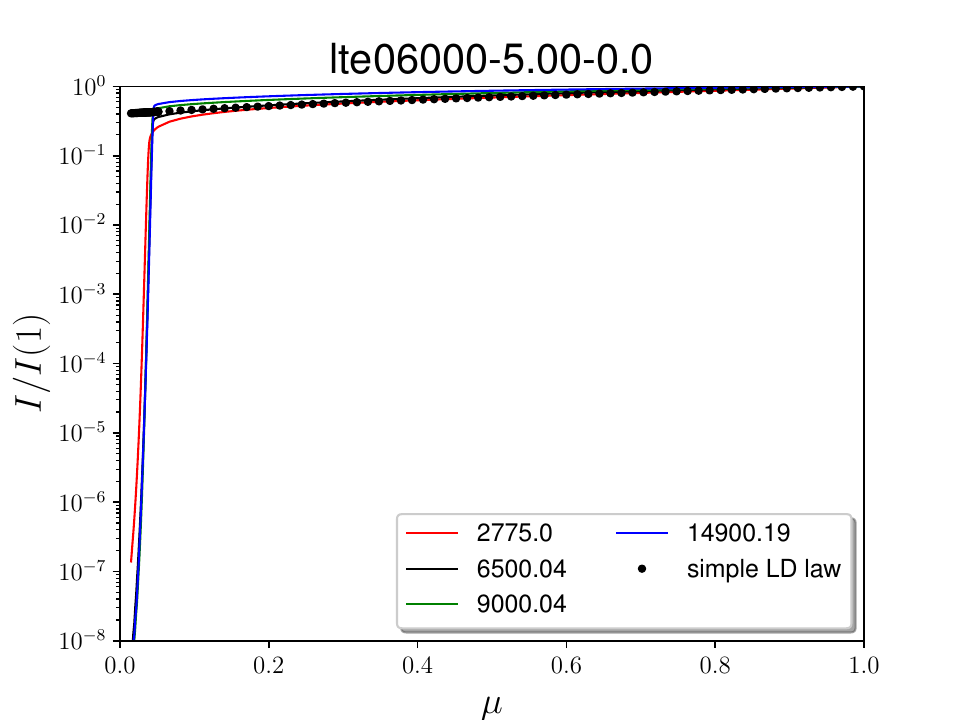}
\end{minipage}
\caption{Limb darkening for representative giant and dwarf models (as indicated in the panels)
for 4 wavelengths. The model with $\Teff=3000\K$, $\logg=0.0$ has a mass of $1.35\Msun$
\AAmarker{and} the one with $\Teff=3000\K$, $\logg=5.0$ has a mass of $0.27\Msun$.
For comparison the simplest plane parallel limb darkening 
law $I(\mu)/I(\mu=1) = \frac{3}{5}\left(\mu+\frac{2}{3}\right)$ is shown as a dotted curve.}
\label{fig:LD}
\end{figure*}

All models in the NewEra grid have been calculated in spherical symmetry (this
is also the case for the previous model grids NextGen, GAIA DR1, and
ACES).
Although for dwarf models the effects of spherical symmetry on the
structure of the atmosphere is small, there are significant effects on the
center-to-limb variation    \citep[limb darkening, see, for example,][]{1966ApJ...143...61C} 
compared to plane parallel
models, as shown in \autoref{fig:LD}.
The model calculations are performed on an adaptive
computational grid generated so that the outer parts of the model atmosphere
are optically thin at all wavelengths.  In plane parallel computations the
intensities at the limb ($\mu=0$)\footnote{$\mu = \cos(\theta)$ where $\theta$
is the angle between the beam of light and the normal to the emitting surface.}
approach a finite value $I(\mu=0)>0$, whereas in spherical
symmetry the intensities drop to zero below a wavelength dependent $\mu$
threshold as the characteristics (the paths of light) closer to the edge of the
star become more and more transparent resulting in large differences compared to plane parallel limb darkening.
The $\mu$ at which the spherical atmosphere becomes transparent (thus setting
the apparent size of the stellar disk) depends on the wavelength due
to the
variation of the opacities with radius. Thus, the apparent size of the model
star is always smaller than the size of the computational grid, that is, 
the last radial grid point.

\begin{figure*}[h]
\centering
    \includegraphics[width=0.49\textwidth]{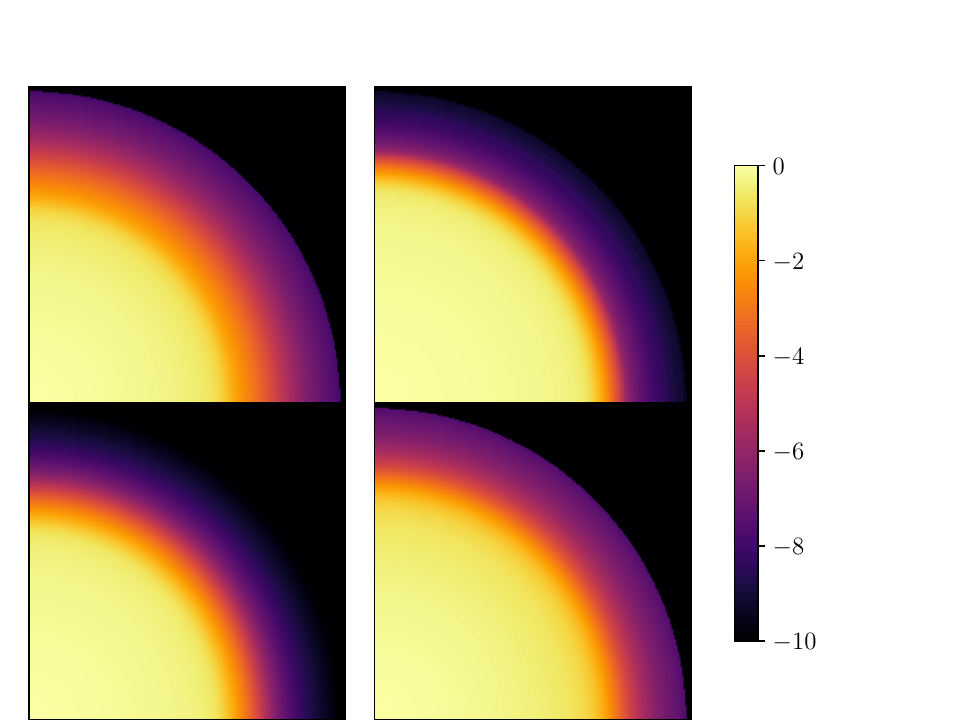}
    \includegraphics[width=0.49\textwidth]{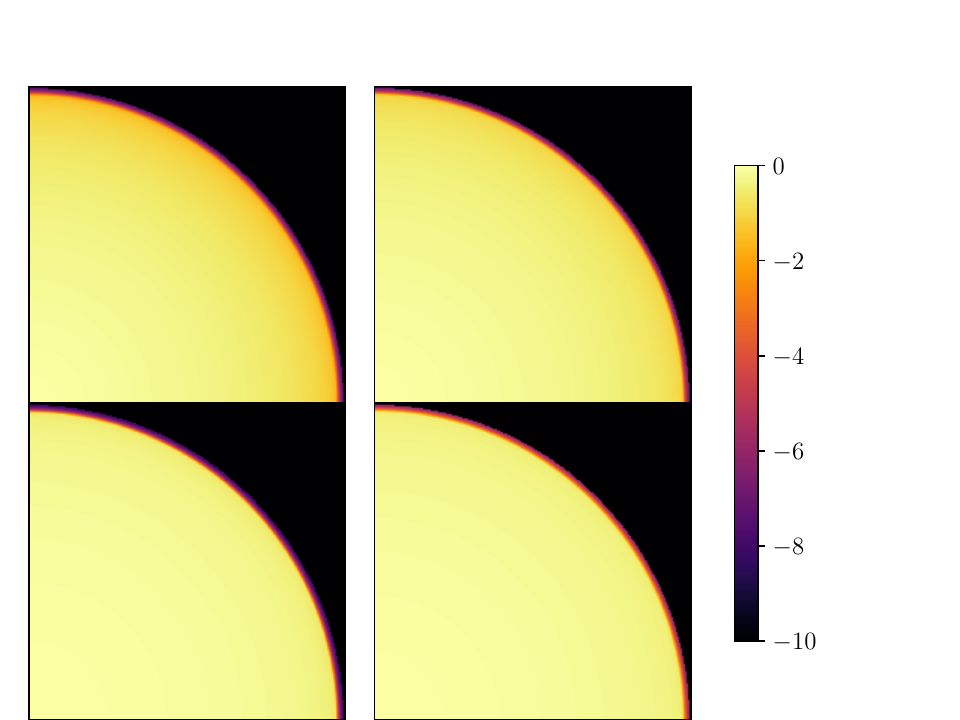}
\caption{\AAmarker{Visualizations of the center-to-limb variation of a
model with $\Teff=3000\K$, $\logg=0.0$ (left panels, $1.35\Msun$) 
and $\logg=5.0$ (right panels, $0.27\Msun$),  solar abundances for 
4 wavelengths: $2775\ang$ (top left), $6500\ang$ (top right),
$9000\ang$ (bottom left), and $1.49\,\mu$m (bottom right). 
The colors indicated by the color bar give $\log_{10}(I/I(\mu=1))$.}
} 
\label{LD2}
\end{figure*}

This effect is much larger and more complex for giants compared to
dwarfs, see 
\Autoref{LD2}. \AAmarker{Dwarfs with a $\logg=5$ have a much smaller 
extension of the atmosphere, causing a much \AAmarker{better-defined}  edge of the star
compared to the giants with $\logg=0.0$ and $0.5$. The larger extension of the giant 
atmospheres causes both a somewhat slower drop off in the outer 
atmosphere and larger electron temperature differences, which in
conjunction with the temperature sensitivity of the line opacities
causes the apparent size of the star to be more wavelength dependent
than is the case for
dwarfs.} However, even for dwarfs the details of the 
center-to-limb variation
are important for interpreting and modeling, e.g., transit light
curves. In applications the \phxO\ 
limb darkening can match the actual limb darkening in the case of transiting systems
\citep[see, for example,][their Extended Data Figure 6]{2014Natur.505...69K}.

\subsection{Data provided}
\label{sec:data_provided}

All model grid and synthetic \AAmarker{spectrum} data are provided as HDF5 (\texttt{.h5}) \AAmarker{\citep{hdf5}} files with 
one file per model. The models are grouped by metallicity [M/H] 
\AAmarker{(for the sake of brevity and filesystem compatibility, Z stands for [M/H] in filenames)} and alpha element variation (if different from zero).
Thus, all models with solar abundances are in a subdirectory \texttt{Z-0.0}, models with metallicity $10^{-4}$ are in the subdirectory
\texttt{Z-4.0} and so on. Correspondingly, the subdirectory \texttt{Z-1.0.alpha=0.2} contains models with an alpha element enhancement of $0.2$\,dex
for a base abundance pattern of [M/H]$=-1.0$ or $1/10$ of the solar metallicity.

Each file follows a naming convention that gives the model parameters $\Teff$, $\logg$, [M/H] and $\alpha$ variation (if $\ne 0$) as well as important code setup details
such as equation of state and model year. The format has the form 
\begin{align*}
    \texttt{lteTTTTTGGGGGZZZZ.alpha=AAAA.PHOENIX}\\ 
    \texttt{-NewEra-ACES-COND-2023.HSR.h5}
\end{align*} 
where lte indicates the LTE
part of the model grid, 
\texttt{TTTTT} is the effective temperature, \texttt{GGGGG} is $-\logg$, \texttt{ZZZZ} the base metallicity and \texttt{AAAA} the $\alpha$ element enhancement,
e.g., 
\begin{align*}
    \texttt{lte04400-5.00-0.0.alpha=-0.2.PHOENIX}\\ 
    \texttt{-NewEra-ACES-COND-2023.HSR.h5} 
\end{align*} 
is a model with $\Teff=4400\K$, $\logg=+5.00$, [M/H]$=0.0$ and $\alpha=-0.2$.
The entry \texttt{PHOENIX-NewEra} describes the model generation and \texttt{ACES-COND} the equation of state (in this case ACES)
with condensation included, \texttt{2023} is the model year and \texttt{HSR} the data product (High Sampling Rate for the wavelengths in the spectrum).

The HSR (high sampling rate) spectra that we provide have an intervals based sampling 
rate\footnote{The sampling distance $\Delta\lambda$ used in the
calculation of our synthetic spectra is \emph{not} to be confused with an
observational resolution element $\Delta\lambda_{\rm obs}$ originating
from observing with a finite aperture.}, $\lambda/\Delta\lambda$, of at least one million
(except from $5.8\,\mu$m to $10\,\mu$m, i.e., between the infrared M and N bands
where the sampling rate is between half a million and one million).
The details are given in \autoref{tab:hsr_params}.
The overall wavelength range (in vacuum) covered is from $900\ang$ to $30\,\mu$m,
resulting in a total  of about $13.1$ million sampled wavelength points. 
The sampling distances $\Delta\lambda$ in the longer wavelength regions
are integer multiples of those in the shorter wavelength regions.
Therefore, it is possible to create evenly spaced spectra across
region borders by dropping data points in the shorter wavelength
regions.

It is important to remember that the sampling rate is not the same
as spectral resolution. In order to compare to a spectrum of a given
resolution, the HSR spectra have to be convolved to the target
resolution with an appropriate filter, e.g., with a Gaussian. For meaningful
results, the target resolution should be less than about 1/10 of the
HSR spectrum sampling rate.

\begin{table}[]
\caption{\label{tab:hsr_params}Wavelength coverage and sampling of the HSR spectra.}
\centering 
\begin{tabular}{r@{ - }lcr@{ - }l}
    \hline
    \hline
    \multicolumn{2}{c}{Range} & $\Delta\lambda$   & \multicolumn{2}{c}{Sampling rate}  \\
    \multicolumn{2}{c}{[$\mu$m]}         &     [m$\ang$]     & \multicolumn{2}{c}{[$10^6$]}   \\
    \hline
       0.09 & 0.32                       &  0.625            & 1.44 & 5.12 \\
       0.32 & 1.37                       &  2.5              & 1.28 & 5.48 \\
       1.37 & 2.64                       &  10               & 1.37 & 2.64 \\
       2.64 & 5.8                        &  20               & 1.32 & 2.9 \\
       5.8  & 30                         &  100              & 0.58 & 3.0 \\
    \hline
\end{tabular}
\end{table}

The HSR spectrum is given in the HDF5 group \texttt{/PHOENIX\_SPECTRUM} in the \texttt{h5} files in the 
following datasets:
\begin{flushright}
\begin{minipage}[c]{0.8\linewidth}
\begin{itemize}
    \item[\texttt{nwl}:] number of wavelength points
    \item[\texttt{wl}:] wavelength array [$\ang$ in vacuum]
    \item[\texttt{flux}:] flux array $\log_{10}(F_\lambda)$ [erg/s/cm$^2$/cm]
    \item[\texttt{bb}:] Planck function array $\log_{10}(B_\lambda(\Teff))$ [erg/s/cm$^2$/cm]
\end{itemize}
\end{minipage}
\end{flushright}

\begin{table}[]
\caption{\label{tab:std_params}Wavelength coverage and sampling of the
standard low sampling rate (LSR) spectra.}
\centering 
\begin{tabular}{r@{ - }lcr@{ - }l}
    \hline
    \hline
    \multicolumn{2}{c}{Range}       & $\Delta\lambda$ & \multicolumn{2}{c}{Sampling rate}  \\
    \multicolumn{2}{c}{[$\mu$m]}    & [$\ang$]        & \multicolumn{2}{c}{}   \\
    \hline
    0.001           &  3            &  0.1            & 100           & $3\times10^5$ \\

    3               &  20           &  1              & $3\times10^4$ & $2\times10^5$ \\
    20              &  50           &  100            & 2000          & 5000          \\
    50              &  600          &  1000           & 500           & 1000          \\
    600             &  1000         &  $10^4$         & 100           & 1000          \\
    \hline
\end{tabular}
\end{table}

For reference a low sampling rate (LSR) spectrum is given in the HDF5 group \texttt{/PHOENIX\_SPECTRUM\_LSR}
in the datasets
\begin{flushright}
\begin{minipage}[c]{0.8\linewidth}
\begin{itemize}
    \item[\texttt{wl}:] wavelength array [$\ang$ in vacuum]
    \item[\texttt{fl}:] flux array $\log_{10}(F_\lambda)$ [erg/s/cm$^2$/cm]
\end{itemize}
\end{minipage}
\end{flushright}
In contrast to the HSR spectra, the LSR spectra cover the whole spectral range from the soft X-ray to the radio wavelength range but at a much lower sampling rate. The details can be found in \autoref{tab:std_params}.

The \phxO\ input Fortran namelist of the calculation of the HSR spectrum from the pre-calculated model 
atmosphere is saved in the dataset \texttt{/PHOENIX\_NAMELIST/phoenix\_nml}.
This is included to quickly extract parameters that are not coded in the 
name of the file, e.g., mass of the star, micro-turbulent speed,  or
the mixing length (using the provided python routine, see \autoref{sec:software})\AAmarker{, however,
most of the contents is not intended for data users but is necessary to recreate the models if needed}.

The HDF5 dataset \texttt{/PHOENIX\_RESTART/phx\_restart} is a string representation of the \phxO\
model restart file used to create the spectrum given in the file so that new spectra can be generated that use the exact same
model atmosphere structure. \AAmarker{The contents of this dataset should be used to extract structure information, such
as radii using the provided python routine, see \autoref{sec:software}. The dataset also includes internal \phxO\ 
data that are only needed for \phxO\ runs and are included here for easier re-running of models with \phxO. The outermost radius  is the
one that should be used when scaling the flux  to that of
a star with a different radius or when calculating the flux of the
model as observed from a given distance.}

The regular (stdout) 
\phxO\ printout for the HSR spectrum \phxO\ run is given as a string in the HDF5 dataset
\texttt{/PHOENIX\_STDOUT/phx\_stdout}. 
This information is 
provided for reference and debugging purposes only.

\section{Summary and conclusions}

We have calculated an LTE grid of stellar models over a wide range of
parameters. The total grid consists of \nmodel spectra different synthetic
spectra in the temperature range $2300\K \le \Teff \le 12000\K$ and $0.0\le
\logg \le 6.0$  and metallicities [M/H] from $-4.0$ to $+0.5$ for all available
$[\alpha/\text{Fe}$]. The number of atomic and molecular lines is an order of
magnitude larger than in our previous grids. We also supply a Python script to
access these data files, which should facilitate their use in studies of, for
example the GAIA data set. Future work will involve inclusion of NLTE effects
into this grid  (P.~Hauschildt et al, in preparation) and extension of this LTE
grid to lower temperatures, applicable to planetary atmospheres (T.~Barman et
al, in preparation).

\section{Data availability \& software products}
\label{sec:software}

In order to facilitate access to 
the model and spectra, we provide a python code \texttt{get\_NewEra\_from\_FDR.py} to download individual models
at \href{https://doi.org/10.25592/uhhfdm.16722}{DOI 10.25592/uhhfdm.16722} \citep{hauschildt_p_h_2025_16723}.
\footnote{at the URL \href{https://doi.org/10.25592/uhhfdm.16722}{https://doi.org/10.25592/uhhfdm.16722}}
The script takes
$\Teff$, $\logg$, [M/H], and the $\alpha$ scale as arguments and downloads the corresponding file.
A list of all available models with download links and MD5 checksums is given in \texttt{list\_of\_available\_NewEra\_models.txt}.

\AAmarker{In addition to the main data product described in the previous section, we 
provide low resolution spectra in a format compatible to the GAIA spectral
library definition. The archive \texttt{NewEra\_for\_GAIA\_DR4.tar} contains the spectra
in the spectral range and resolution required for the GAIA mission, whereas 
the archive \texttt{PHOENIX-NewEra-LowRes-SPECTRA.tar.gz} contains the 
spectra from $2500\,\ang$ to $2.5\,\mu$m with a resolution (Gaussian filter standard deviation) and sampling rate of 
$0.1\ang$, and finally the archive \texttt{PHOENIX-NewEra-JWST-SPECTRA.tar.gz}
gives the spectra in the wavelength range and resolution useful for JWST: between
$0.6\,\mu$m and $28.5\,\mu$m with a resolution \AAmarker{(Gaussian filter standard deviation)} and sampling rate of $2\ang$. Note
that the wavelengths and fluxes in these files are given in nm and W/m$^2$/nm, respectively, in
order to be compatible to the GAIA spectral library. Each archive contains 
the spectra for each abundance pattern in individual files and were
generated for the HSR spectra by convolution with a Gaussian filter.
\texttt{example\_read\_gaia\_fmt.py} is a simple example python code to 
read the first spectrum of an individual archive file.}

We also provide the Python routine \texttt{example\_read\_HSR\_H5.py} to access the contents
of the provided h5 files. The 
routine reads the supplied HDF5 files and returns the data as strings, scalars and arrays. 
This includes examples of the stored metadata for the
particular grid entry and additional output information that can be
accessed. \AAmarker{The python routine \texttt{example\_read\_structure\_from\_HSR\_H5.py} shows
how to access the structure data in the \texttt{/PHOENIX\_RESTART/phx\_restart} dataset, in particular 
radii, gas and electron pressures, and electron temperature.} 

\AAmarker{Due to size restrictions, limb darkening data need to be generated and 
are made available upon request.}

\begin{acknowledgements}
Data processing in this work was facilitated by GNU parallel \citep{tange_2022_7105792}.
The model grid calculations presented here were partially performed at the National Energy
Research Supercomputer Center (NERSC), which is supported by the Office of
Science of the U.S.  Department of Energy under Contract No. DE-AC03-76SF00098.
The authors gratefully acknowledge the computing time made available to them on the high-performance computers HLRN-IV at GWDG  at the NHR Center NHR@Göttingen and at ZIB at the NHR Center NHR@Berlin. These Centers are jointly supported by the Federal Ministry of Education and Research and the state governments participating in the NHR (www.nhr-verein.de/unsere-partner). 
Additional computing time was provided by the RRZ computing clusters Hummel and Hummel2.
We thank all these institutions for a generous allocation of computer time.
PHH and EB are grateful for a NERSC Science Acceleration Program
(NESAP) awards from NERSC (2014--2020) which 
allowed for the efficient use of ``Burst Buffer'' technology. 
PHH gratefully acknowledges the support of NVIDIA Corporation with the donation
of a Quadro P6000 GPU used in this research.
TB acknowledges support by NASA-XRP grant 80NSSC21K0572.
EB acknowledges support by NASA grants
JWST-GO-2114, JWST-GO-2124, JWST-GO-4436,  JWST-GO-4522, JWST-GO-3726,
JWST-GO-4217, JWST-GO-5057, JWST-GO-5290, JWST-GO-6023, JWST-GO-6582,
HST-AR-17555, and 80NSSC20K0538. 
Support for programs \#2114, \#2124, \#4436,  \#4522, \#3726, \#4217,
\#5057, \#5290, \#6023, \#6582, and \#17555 were 
provided by NASA through a grant from the Space Telescope Science 
Institute, which is operated by the Association of Universities for 
Research in Astronomy, Inc., under NASA contract NAS 5-03127.
\AAmarker{AS acknowledges support by the DFG grant SCHW 1358/5-1 within the priority program SPP 1992 ''Exploring the Diversity of Extrasolar Planets''.}
\end{acknowledgements}

\bibliography{NewEra_bib}

\end{document}